\RequirePackage{fix-cm}
\documentclass[natbib]{svjour3}

\smartqed  

\usepackage{amssymb}
\usepackage{siunitx}
\usepackage{natbib}        
\usepackage{amsmath}
\usepackage{graphicx}
\usepackage[T1]{fontenc}
\usepackage[utf8,latin1]{inputenc}
\usepackage{epsfig}
\usepackage{txfonts}
\usepackage{url}
\usepackage{color}
\usepackage{xcolor}

\usepackage{bm}
\usepackage{sgl-commands}
\usepackage[normalem]{ulem}

\usepackage{geometry}
\newcommand{\btheta}{\bm{\theta}}
\newcommand{\balpha}{\bm{\alpha}}

\newcommand{\chapintro}{\refedit{\citet{bib3:Saha24}}}
\newcommand{\chapintroalt}{\refedit{\citealt{bib3:Saha24}}}

\newcommand{\chapho}{\citet{bib3:Birrer24}}
\newcommand{\chaphoalt}{\citealt{bib3:Birrer24}}

\newcommand{\chapdm}{\refedit{\citet{bib3:Vegetti23}}}
\newcommand{\chapdmalt}{\refedit{\citealt{bib3:Vegetti23}}}

\newcommand{\chapmicro}{\refedit{\citet{bib3:Vernardos24}}}
\newcommand{\chapmicroalt}{\refedit{\citealt{bib3:Vernardos24}}}

\newcommand{\chapsearch}{\refedit{\citet{bib3:Lemon24}}}

\newcommand{\thischapter}{review article}

\newcommand{\refedit}[1]{{{#1}}}

\journalname{my journal}

\begin{document}

\title{Strong Lensing by Galaxies}

\author{A.~J.~Shajib$^{1, 2,\dagger}$\thanks{$\dagger$ NHFP Einstein Fellow}, G.~Vernardos$^{3,4,5,\star}$\thanks{$\star$ georgios.vernardos@epfl.ch}, T.~E.~Collett$^{6}$, V.~Motta$^{7}$, D.~Sluse$^{8}$, L.~L.~R.~Williams$^{9}$, P.~Saha$^{10}$, S.~Birrer$^{11,12,13}$, C.~Spiniello$^{14, 15}$, T.~Treu$^{16}$}

\authorrunning{A.~J.~Shajib et al.}

\institute{ 
    1. Department of Astronomy and Astrophysics, 
    University of Chicago,
    Chicago, IL 60637, USA \\ \email{ajshajib@uchicago.edu} \\
    2. Kavli Institute for Cosmological Physics,
    University of Chicago,
    Chicago, IL 60637, USA \\
    3. Institute of Physics, Laboratory of Astrophysics,
    Ecole Polytechnique F\'ed\'erale de Lausanne (EPFL), 
    Observatoire de Sauverny, 1290 Versoix, Switzerland \\
    4. Department of Astrophysics, American Museum of Natural History, Central Park West and 79th Street, NY 10024-5192, USA \\
    5. Department of Physics and Astronomy, Lehman College of the CUNY, Bronx, NY 10468, USA\\
    6. Institute of Cosmology and Gravitation, 
    University of Portsmouth,
    Burnaby Rd, Portsmouth PO1 3FX, UK \\
    7. Instituto de Fisica y Astronomia, Facultad de Ciencias, 
    Universidad de Valparaiso,
    Avda. Gran Bretana 1111, Valparaiso, Chile \\
    8. STAR Institute,
    Quartier Agora - All\'ee du six Aout,
    19c B-4000 Li\'ege, Belgium \\
    9. School of Physics and Astronomy,
    University of Minnesota,
    116 Church Street SE, Minneapolis, MN 55455, USA \\
    10. Physik-Institut, University of Zurich,
    Winterthurerstrasse 190, 8057 Zurich, Switzerland \\
    11. Kavli Institute for Particle Astrophysics and Cosmology and Department of Physics,
    Stanford University,
    Stanford, CA 94305, USA \\
    12. SLAC National Accelerator Laboratory,
    Menlo Park, CA 94025, USA \\
    13. Department of Physics and Astronomy,
    Stony Brook University,
    Stony Brook, NY 11794, USA \\
    14. Department of Physics,
    University of Oxford,
    Denys Wilkinson Building, Keble Road, Oxford OX1 3RH, UK \\
    15. INAF -- Osservatorio Astronomico di Capodimonte,
    Via Moiariello 16, 80131, Naples, Italy \\
    16. Department of Physics and Astronomy,
    University of California, Los Angeles,
    430 Portola Plaza, Los Angeles, CA 90095, USA
}

\date{Received: date / Accepted: date}

\maketitle

\begin{abstract}
Strong gravitational lensing at the galaxy scale is a valuable tool for various applications in astrophysics and cosmology. Some of the primary uses of galaxy-scale lensing are to study elliptical galaxies' mass structure and evolution, constrain the stellar initial mass function, and measure cosmological parameters. Since the discovery of the first galaxy-scale lens in the 1980s, this field has made significant advancements in data quality and modeling techniques. In this review, we describe the most common methods for modeling lensing observables, especially imaging data, as they are the most accessible and informative source of lensing observables. We then summarize the primary findings from the literature on the astrophysical and cosmological applications of galaxy-scale lenses. We also discuss the current limitations of the data and methodologies and provide an outlook on the expected improvements in both areas in the near future.

\end{abstract}

\label{chapter3}

\section{Introduction} \label{sec3:introduction}

This \thischapter\ discusses applications of galaxy-scale strong lenses to study the properties of the deflector galaxies, that is, the central lensing galaxies. These have so far typically been massive elliptical galaxies at $0.1 \lesssim z \lesssim 1$ and strong lensing has been chiefly applied to study their internal structure and composition. However, a large sample of strong lenses with deflectors other than massive ellipticals is expected to be discovered in this decade from the upcoming deep sky surveys. We can also gain important insights into the formation and evolution of the deflectors by comparing their structural properties (e.g., the logarithmic slope of the density profile and the dark matter fraction) across cosmic times. Furthermore, in this \thischapter, we present cosmological applications of the galaxy-scale lenses that do not require time delay information -- that is, measuring cosmological parameters such as the matter density parameter $\Omegam$ and the dark energy equation-of-state parameter $w_{\rm de}$. Cosmological application involving the time delay measurements, that is, measuring primarily the Hubble constant, is reviewed by \chapho. Additionally, \chapdm\ review the application of galaxy-scale lensing to study sub-galactic structures of dark matter.

In this introductory section, we provide a brief description of the lensing phenomenology (\secref{sec3:lensing_phenomenology}) and discuss the advantages of strong-lensing observables in comparison with other probes of galaxy mass, such as stellar dynamics (\secref{sec3:lensing_advantages}).
The remainder of this \thischapter\ is organized as follows. In \secref{sec3:historical_results}, we highlight the significant historical results involving galaxy-scale lenses and introduce several prominent lens samples. In \secref{sec3:observables_and_methods}, we describe the strong-lensing observables and their modeling and analysis methods. We discuss the application of galaxy-scale strong lensing to study galaxy properties and evolution in \secref{sec3:science_applications} and to constrain cosmological parameters in \secref{sec3:cosmology}. Next, in \secref{sec3:open_problems}, we discuss open issues -- both in technical aspects and scientific questions -- and provide future outlooks. We conclude the \thischapter\ in \secref{sec3:conclusion}.

\subsection{Brief description of lensing phenomenology at the galaxy-scale} \label{sec3:lensing_phenomenology}

\begin{figure*}
	\includegraphics[width=\textwidth]{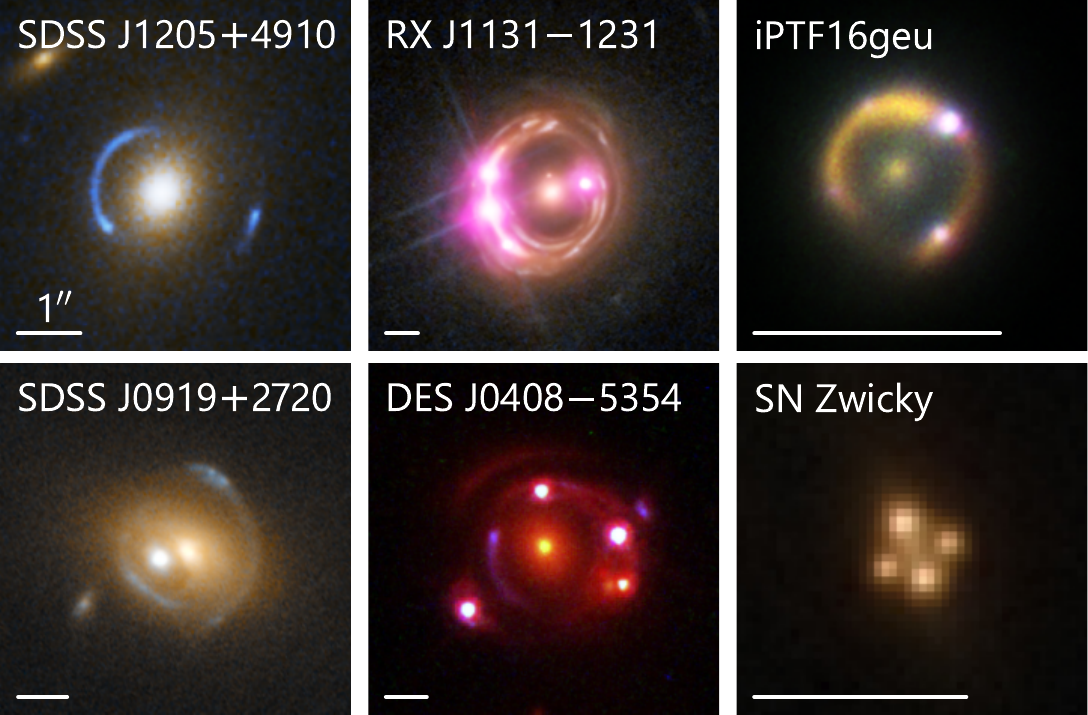}
	\caption{\label{fig3:lens_montage}
	Examples of galaxy-scale lenses with different types of background sources. These are false color images created from multi-band \textit{HST} imaging, and in some cases, also combined with the \textit{Chandra} X-ray data (RX J1131$-$1231) and the Keck Observatory IR imaging (iPTF16geu). The first column shows two lenses with background galaxies without any resolved point source \citep{bib3:Bolton06, bib3:Courbin12}. The second column shows two lensed quasar systems \citep{bib3:Suyu13, bib3:Shajib20}. The third column shows two lensed supernovae \citep{bib3:Goobar17, bib3:Pierel23}. The white bar in each panel represents 1$\arcsec$. \textit{Image credits: NASA, ESA, A.~Bolton, the SLACS team, {\rm Chandra}, A.~J.~Shajib, W.~M.~Keck Observatory, T.~Li, and J.~Pierel}.
 }
\end{figure*}

The background extended source in a galaxy-scale strong lens can be lensed into multiple arcs or a complete Einstein ring. Multiple point images will also appear if there is a point source within the background galaxy, for example, an active galactic nucleus (AGN) or quasar, or a supernova (see \figref{fig3:lens_montage}). The former type of system is called galaxy--galaxy lenses. In contrast, the latter is usually referred to as `quads' (for the case of 4 detected point images) or `doubles' (for the case of 2 detected point images). The different manifestations of strong lenses -- the appearance of arcs or a full Einstein ring, or the number of point images -- depend on the position of the source with respect to the lens caustics \citep[as introduced in][]{bib3:Saha24}. Strong lensing of point sources can provide three types of observables: image positions, image magnification ratios, and time delays between the images. It follows from the lensing theory that all three are properties of the Fermat potential or the arrival-time surface. Images form at the local extrema -- minima, saddle points, and maxima -- of this potential. Magnification is inversely proportional to the determinant of the Fermat potential's Jacobian matrix. Lastly, time delays are the differences in the Fermat potential at the image locations (for a detailed explanation, see \chapintroalt).

However, not all of these observables are available for every lens. Image positions are almost always observable for point-source lens systems. While always observable, image flux ratios can only be turned into actual magnifications when the intrinsic source flux is known. However, interpreting these magnifications during modeling requires extra care, as they can be affected by more complex features in the galaxy-scale mass distribution, for example, baryonic disks \citep{bib3:Hsueh16, bib3:Hsueh17}, microlensing by individual stars and planets in the lensing galaxy (see \chapmicroalt), intermediate-mass-scale structures like dark matter subhalos (see \chapdmalt), or dust extinction by the lens galaxy \citep[e.g.,][]{bib3:Motta02, bib3:Mediavilla05}.
Time delays can be obtained only for variable point sources, like quasars, supernovae, and, in the future, gravitational waves and fast radio bursts (see \chaphoalt). Measuring time delays can be a difficult task, especially for quasars, which require long-term monitoring spanning from a few seasons to years \citep[e.g.,][]{bib3:Eigenbrod05, bib3:Bonvin16, bib3:Millon20}.

The above description of point-source lensing must be slightly modified for extended sources comparable in size to the lens caustics. Instead of point-like image positions,
the light from such a source is spread into an extended area in the image plane and then further smeared by the point spread function.
As a result, it is hard to know \textit{a priori} how the observed flux of the multiple images of the source on the lens plane traces back to the same location on the source plane.
Therefore, even if the actual source brightness is known, a lens model is required to compute the magnification field (i.e., the Jacobian of the lens potential) across the lens plane.
Due to their size, extended sources are not variable on time scales relevant for lensing. Thus, time delays are not observable without a variable point source. Note that extended arcs from the point source's host galaxy are also usually present for a lens system that includes a point source. Therefore, we can constrain the lens model by simultaneously utilizing the lensed arcs' flux distribution and the point-image positions.

\subsection{Unique advantages of lensing as a probe of galaxy structure} \label{sec3:lensing_advantages}

Other than lensing, the only commonly used probe of the mass distribution in galaxies is kinematics, that is, the velocity dispersion or streaming motions of stars, gas, or globular clusters. For this to be fully informative of the mass distribution, however, spatially resolved measurements are necessary \citep{bib3:Cappellari16}, which is often limited to nearby galaxies at $z \le 0.5$ due to the sizeable observational cost, or due to the smaller size on the sky of objects at higher redshifts. Using individual stars' motions, for example, from \textit{Gaia}, to map a galaxy's mass is only limited to our Milky Way \citep[e.g.,][]{bib3:Nitschai20}. Thus, an aperture-integrated velocity dispersion measurement \refedit{(from either long-slit or integral field spectroscopy)} is the only possible dynamical observable for galaxies other than ours. However, strong-lensing observables are obtained from high-resolution imaging data, which are much more informative than the dynamical observables for galaxies beyond the local Universe \citep[$z \gtrsim 0.03$, e.g.,][]{bib3:Smith15}. \refedit{This is especially advantageous for studying the lens galaxies since the lensing signal does not depend on the surface brightness of the galaxy being studied, unlike stellar kinematics, but on the combination of the lens galaxy mass and the source brightness.} Strong lensing can provide $\sim$1--2\% constraints on the mass enclosed within the Einstein radius from imaging data alone. To obtain a similarly precise mass constraint from dynamical observations at high redshift, either integration times longer by a factor $\mathcal{O}(10)$ are needed on current facilities, or we must wait for better quality adaptive optics systems planned for future extremely large telescopes.

Furthermore, the dynamical observables have their own intrinsic degeneracies, that is, the mass--anisotropy degeneracy %
\citep[e.g.,][]{bib3:Treu02b}. There is an important complementarity between lensing and dynamical observables, which can be used to break their corresponding degeneracies \citep[e.g.,][]{bib3:Courteau14}. Finally, gravitational lensing responds to both baryonic and dark matter without any assumptions on their dynamical state, regardless of whether it is in equilibrium or not.

\section{Historical background} \label{sec3:historical_results}

This section provides a historical note on the initial discoveries of galaxy-scale strong lenses (\secref{sec3:initial_discoveries}). Then, in \secref{sec3:lens_samples}, we briefly introduce several prominent samples of galaxy-scale lenses that have contributed to the major science applications described in \secref{sec3:science_applications} and \secref{sec3:cosmology}. Additional references can be found in the review by \citet{bib3:Treu10b}.

\subsection{Initial discoveries of strong lensing systems} \label{sec3:initial_discoveries}

The first strong lens, the double quasar 0957$+$561A,B, was discovered in 1979 by \cite{bib3:wal79}. The system consists of two images of the quasar separated by $5\,\arcsecf{7}$. The authors first offered a `conventional' interpretation that the two images are different, individual quasars that happen to be close to each other and share the same physical characteristics.
Since no gravitational lens was known before that, their less conventional view was that the two are multiple images of the same source.
Subsequent work showed that the lensing hypothesis was correct. For example, the structure of the radio jets emanating from the two images of the quasar is consistent with them being mirror imaged, as one would expect for a minimum and a saddle-point image \citep{bib3:Gorenstein88, bib3:Garrett94}. 
The discovery of the first quadruply imaged quasar, PG 1115$+$080, was announced the following year \citep{bib3:Weymann80}. It was initially called a `triple' because the second arriving minimum and its neighboring saddle point were too close to be resolved. These two lenses' discoveries opened up a new field in astrophysics: observations of multiply-imaged, `strong' gravitational lenses.  Discoveries of other types of lensed sources followed: for example, the first dust-obscured Seyfert 2 AGN, IRAS F10214$+$4724, was detected in the infrared (IR), and later identified as being lensed \citep{bib3:Eisenhardt96, bib3:Lehar96}.

The increasing number of detections of such lensed systems spurred a lens modeling effort. The initial studies that used more detailed models beyond a point-like lens mass appeared in the early 1980s \citep{bib3:Young80, bib3:Young81a, bib3:Young81b}, and already recognized ``that there are several plausible ways to reproduce the observations'', foreshadowing the importance of lens model degeneracies. The role of the lens mass granularity due to individual stars in the lensing galaxy was also recognized very early on \citep{bib3:Chang79, bib3:Young81c} and later grew into the rich sub-field of extragalactic stellar microlensing (see \chapmicroalt\ for a review).

\subsection{Prominent samples of galaxy-scale lenses} \label{sec3:lens_samples}

In this subsection, we briefly introduce some of the most prominent samples of galaxy-scale lenses that had an impact on the science applications presented in \secref{sec3:science_applications} and \secref{sec3:cosmology}. Note that this is not a complete list of all the discovered galaxy-scale lenses. The specifics of lens searching and discovery to build samples like these are reviewed by \chapsearch, which we refer the reader to for recent developments in the search methods and discoveries.

\subsubsection{MG-VLA survey-based samples}

The first systematic search for strongly lensed systems at radio wavelengths took place in the eighties within the \refedit{MIT--Greenbank--Very Large Array (MG-VLA)} survey \citep{bib3:Lawrence86}. This survey discovered a few famous radio-loud lensed quasars among thousands of radio sources scrutinized with high resolution by the Very Large Array (VLA). The Jodrell Bank--VLA Astromtric Survey \citep[JVAS;][]{bib3:Patnaik92, bib3:King96}, and its successor, the Cosmic Lens All Sky Survey \citep[CLASS;][]{bib3:Myers95} was the largest survey carried out for a long time. This survey targeted the whole northern sky ($0^{\circ} < \textrm{Dec} < 75^{\circ}$) for multiple images (separated by $0\,\arcsecf{3} < \Delta \theta < 6\,\arcsecf{0}$) among flat spectrum radio sources brighter than 30\,mJy. CLASS discovered 22 new systems, among which twelve are doubles, nine are quads, and one displays six images \citep{bib3:Browne03}.

\subsubsection{The CASTLES sample}

The CfA-Arizona Space Telescope LEns Survey (CASTLES\footnote{\url{https://lweb.cfa.harvard.edu/castles/}}) is a follow-up \textit{Hubble Space Telescope} (\textit{HST}) imaging survey of $\sim$100 galaxy-scale lenses\footnote{There are a handful cluster-scale lenses included in this sample as well.} known at the time, some from previous surveys such as CLASS \citep{bib3:Munoz98} \refedit{and others from serendipity or targeted surveys \citep[for details, see][]{bib3:Lemon24}.} This survey collected the first uniform ensemble of high-resolution images of known galaxy-scale lens systems, including both galaxy--quasar and galaxy--galaxy lenses. An account of early systematic lens searches, which generally unveiled small samples of less than six systems, can be found in \citet{bib3:Claeskens02}.

\subsubsection{Samples of lensed sub-mm galaxies}

\refedit{Lens searches in the sub-mm have proved to be efficient in finding hundreds of lensed dusty star-forming galaxies at high redshift ($z \sim 1$--$4$) with high purity in the candidate sample, using the sharp cutoff in the luminosity function for these galaxies \citep[see][]{bib3:Lemon24}. The initial samples detected using this technique came from the \textit{Herschel} Astrophysical Terahertz Large Area Survey (HATLAS) and the \textit{Herschel} Multi-tiered Extragalactic Survey \citep[HerMES;][]{bib3:Negrello10, bib3:Negrello16}. Since these samples of lenses are source-selected, the selection function of the lens galaxies is less affected than for lens-selected samples, thus providing an advantageous avenue to study galaxy properties \citep[e.g.,][]{bib3:Dye14, bib3:Dye18, bib3:Amvrosiadis18, bib3:Maresca22}.}

\subsubsection{SDSS-based samples}

The Sloan Lens ACS (SLACS) survey discovered 85 galaxy--galaxy lenses from the Sloan Digital Sky Survey (SDSS) spectroscopic data \refedit{by identifying multiple redshifts in the fiber (with 3$\arcsec$ diameter) spectra.} The SLACS survey also followed these systems up with multi-band \textit{HST} imaging \citep{bib3:Bolton06, bib3:Auger09}. The sample was expanded with 40 new systems with smaller deflector masses by the SLACS for the Masses (S4TM) sample \citep{bib3:Shu15}. In addition to galaxy--galaxy lenses, the SDSS Quasar Lens Search (SQLS) discovered a sample of 28 galaxy--quasar lenses using SDSS multicolor imaging data \citep{bib3:Oguri06}. More galaxy--quasar lens systems were discovered from joint SDSS and UKIRT Infrared Deep Sky Survey (UKIDSS) data by the Major UKIDSS--SDSS Cosmic Lens Survey \citep[MUSCLES;][]{bib3:Jackson12}.

\subsubsection{CFHTLS-based samples}

The Strong Lensing Legacy Survey \citep[SL2S;][]{{bib3:Gavazzi12}} discovered a sample of $\sim$35 galaxy--galaxy lenses from the Canada--France--Hawaii Telescope Legacy Survey (CFHTLS) data. Some newer lens samples have also been discovered from this survey \citep{bib3:More16, bib3:Paraficz16}.

\subsubsection{BOSS-based samples}

The Baryon Oscillation Spectroscopic Survey Emission-Line Lens Survey (BELLS) discovered $\sim$30 galaxy--galaxy systems from the Baryon Oscillation Spectroscopic Survey (BOSS) and obtained \textit{HST} imaging for them \citep{bib3:Brownstein12}. This survey was later expanded into the BELLS for the GALaxy-Ly$\alpha$ EmitteR sYstems (BELLS GALLERY) survey, where the source galaxies are specifically Ly$\alpha$ emitters \citep{bib3:Shu16}. A sample of 13 strongly lensed quasars has also been discovered from the BOSS data \citep{bib3:More16b}.

\subsection{Other samples and ongoing efforts from recent surveys}

Numerous large-area sky surveys have recently discovered several other lens samples. The STRong-Lensing Insights into the Dark Energy Survey (STRIDES) collaboration has discovered $\sim$30 quadruply lensed quasar systems from the Dark Energy Survey (DES) data -- often in combination with data from other sky surveys -- and obtained multi-band \textit{HST} imaging of them in IR, optical, and ultra-violet (UV) bands \citep{bib3:Shajib19, bib3:Schmidt23}. Many galaxy--galaxy lens candidates have also been identified in the DES data \citep{bib3:Jacobs19, bib3:Jacobs19b, bib3:Rojas21, bib3:Tran22}. In addition to the DES, surveys such as the Canada--France Imaging Survey (CFIS), \textit{Gaia}, the Hyper Suprime-Cam (HSC) survey, the Kilo Degree Survey (KiDS), and the Panoramic Survey Telescope and Rapid Response System (Pan-STARRS) have provided a plethora of newly discovered galaxy-scale strong lenses \citep[e.g.,][]{bib3:Petrillo17,bib3:Petrillo19,bib3:Agnello18, bib3:KroneMartins18, bib3:Delchambre18, bib3:Lemon18, bib3:Lemon20, bib3:Canameras21, bib3:Savary21, bib3:Li21, bib3:Lemon22, bib3:Wong22}. Most of these samples still contain candidate lenses and require spectroscopic confirmation (by measuring the redshifts) and high-resolution imaging (to perform lens modeling) for the science applications described in \secref{sec3:science_applications} and \secref{sec3:cosmology}. See \chapsearch\ for a detailed discussion of the recent and ongoing efforts.

\section{Observables and analysis methods} \label{sec3:observables_and_methods}

This section describes the strong-lensing observables (\secref{sec3:lensing_observables}) and the analysis techniques to constrain galaxy properties from them. We describe the lens modeling methods in \secref{sec3:modeling_methods} and the commonly used models in \secref{sec3:lens_mass_models}. We outline the Bayesian hierarchical framework in \secref{sec3:hierarchical_framework}, which allows inferring population characteristics of galaxies from a sample of lenses. Lastly, non-lensing observables most commonly combined with strong lensing ones are presented in \secref{sec3:non_lensing_observables}.

\subsection{Lensing observables} \label{sec3:lensing_observables}

The two types of lensing observables for galaxy-scale lenses are imaging of the lensing system (\secref{sec3:data_imaging}) and the time delay between a pair of images from a point source (\secref{sec3:data_time_delay}).

\subsubsection{Imaging of the lens system} \label{sec3:data_imaging}
The most common and informative lensing observables result from imaging data with angular resolution much better than the Einstein radius.
An extended source can be lensed into clearly identifiable arcs that can form partial or complete Einstein rings (see \figref{fig3:lens_montage}).
The conjugate points on the arcs, that is, locations that are traced back to the same location in the source plane, can be used to simultaneously constrain a lens model (through its deflection angles and magnification) and reconstruct the surface brightness of the source that is \textit{a priori} unknown.
Although simple models can be constrained from any image that displays lensing features with sufficient signal-to-noise ratio, high-resolution imaging from space- or ground-based telescopes offers many more observational constraints (i.e., conjugate pixels).
This is crucial for exploring more sophisticated models, which are required for precise science applications (e.g., see \secref{sec3:science_applications} and \secref{sec3:cosmology}).

If the background source contains a point-like emitting region -- for example, a quasar or a supernova -- the positions of its multiple images (i.e., conjugate points) can be extracted from the data and used as constraints for a lens model.
Although the simple addition of two or four conjugate points may initially seem insignificant compared to the hundreds of pixels that correspond to the extended source, the flux contained in the corresponding pixels may be brighter than the entire host galaxy, for example, in the case of an AGN, \refedit{and their positions can be constrained with sub-pixel accuracy \citep[e.g.,][]{bib3:Shajib19, bib3:Schmidt23}}. Hence, these conjugate points have a strong effect on the resulting model and must be treated separately. In addition to the astrometry of the point-like images, their magnification ratio (or flux ratio) can be used as an observational constraint. However, image magnifications are susceptible to micro- and milli-lensing, dust extinction\footnote{Lensed arcs from extended sources are also vulnerable to dust extinction.}, and the effect of higher order moments in the mass distribution of the lens usually attributed to the complex, non-linear physics of baryons, for example, galactic disks and bars \citep{bib3:Hsueh16, bib3:Hsueh17}.
Therefore, many lensing analyses involving systems with %
such point-like sources choose not to use magnification ratios as constraints \citep[e.g.,][]{bib3:Shajib19}. However, suppose the microlensing and dust extinction effects can be incorporated and quantified within a lens model. In that case, any residual flux ratio anomaly would signal a departure from the smooth macro-model for the deflector galaxy and thus could indicate the presence of sub-galactic dark matter structures within the galaxy-scale halo \citep[e.g.,][]{bib3:Mao98, bib3:Nierenberg17, bib3:Gilman20}. Such detection of dark substructures can provide important insights into the nature of the dark matter, as described in \chapdm.

\refedit{Lens modeling by simultaneously fitting the imaging data in multiple bands (from radio wavelengths to UV) has become commonly employed in the literature when such data are available \citep[e.g.,][]{bib3:Dye14, bib3:Oldham18, bib3:Shajib19, bib3:Young22, bib3:Tan24}.
\refedit{Such multi-band modeling has the advantage of adding constraints in regions of the images that may be poorly detected in some wavelength ranges and deblending the lensing galaxy from the images and arcs. The main drawback is the increase in model complexity due to the wavelength dependence of the lens galaxy and source morphology.} However, such multi-band lens modeling is expected to be ubiquitous in the upcoming decade, with multi-band data being more available thanks to the current and upcoming facilities such as the \textit{JWST} and the Vera Rubin Observatory \citep{bib3:Shajib24}. \refedit{Several automated modeling pipelines (for both single-band and multi-band data) are being developed to tackle the computational aspect of modeling very large lens samples \citep{bib3:Oguri10, bib3:Collett15} to be discovered by the Rubin Observatory, \textit{Euclid}, and the \textit{Roman Space Telescope} \citep[e.g.,][]{bib3:Chan15, bib3:Nightingale18, bib3:Shajib21, bib3:Etherington22, bib3:Schmidt23, bib3:Tan24}.}
}

\subsubsection{Time delay} \label{sec3:data_time_delay}
If the background source is a variable point source, for example, a quasar or a supernova, the delay between the arrival times of photons at its multiple different images can be measured through long-term monitoring that spans from a few months to decades \citep[e.g.,][]{bib3:Eigenbrod05, bib3:Bonvin16, bib3:Millon20}. The time delays are variant under the well-known mass-sheet degeneracy (MSD) in lensing, unlike the imaging observables \citep{bib3:Falco85, bib3:Schneider14}. Thus, they can be combined with the imaging observables to break the MSD when constraining the potential of the lensing galaxy. However, such a combination of these observables requires a fiducial cosmological model since the time delays depend on the cosmology, particularly the Hubble constant. A more detailed discussion on the measurement of time delays is provided in \chapho.

\subsection{Lens modeling methods} \label{sec3:modeling_methods}
Lens modeling is the process of constraining properties of the lens galaxy and the source from the lensing observables.
Traditional methods are based on reconstructing the source light and lens potential to fit the data under some assumptions, such as regularization, and are described in \secref{sec3:likelihood_based_inference}.
More recently, machine learning methods are being developed for these purposes, which we present in \secref{sec3:machine_learning_methods}. \refedit{It is beyond the scope of this review to provide a beginner's guide to modeling galaxy-scale lenses, but we refer the reader to \citet{bib3:Saha24} for an introduction to lens modeling. Readers interested in the science applications without needing a technical discussion on lens modeling and analysis techniques may go directly to \secref{sec3:science_applications} and refer back to the rest of this \secref{sec3:observables_and_methods} as needed.}

\subsubsection{Likelihood-based inference} \label{sec3:likelihood_based_inference}

Such methods require a likelihood function that leads to an optimized model that can reproduce an observed multiply-imaged system down to the noise level. \refedit{We note that whereas an optimized forward model aims to reproduce the data to the noise level, it is often difficult to achieve that in practice, often owing to the simplifying assumptions made in the model. As a result, fine-tuning the model complexity is commonly required through trial-and-error to meet the accuracy requirement for a given science case \citep[e.g.,][]{bib3:Shajib19, bib3:Schmidt23}.}
In general, the \refedit{lens} model consists of three main components: the background source's flux distribution, the mass distribution in the lensing galaxy (or galaxies), and the flux distribution in the lensing galaxy (or galaxies). %
The likelihood function that measures the goodness-of-fit of the model to the data can be defined as
\begin{equation} \label{eq3:flux_likelihood}
	\mathcal{L}(\boldsymbol{d} \mid \boldsymbol{m}) \propto \exp \left[ -\frac{1}{2}\left(\boldsymbol{d} - \boldsymbol{m}\right)^{T} \mathsf{\Sigma}_{\boldsymbol{d}}^{-1}\left(\boldsymbol{d} - \boldsymbol{m}\right) \right],
\end{equation}
where $\boldsymbol{d}$ is the vector of pixel values in the data and 
\begin{equation}
    \boldsymbol{m} \equiv \boldsymbol{m}(\boldsymbol{\xi}_{\mathrm{mass}},\  \boldsymbol{\xi}_{\mathrm{source}},\ \boldsymbol{\xi}_{\mathrm{light}})
\end{equation} 
is the model-computed flux in the data pixels. Here, $\boldsymbol{\xi}_{\rm mass}$ is the set of model parameters defining the lens mass distribution, $\boldsymbol{\xi}_{\rm source}$ is the set of model parameters defining the source's flux distribution, $\boldsymbol{\xi}_{\rm light}$ is the set of model parameters defining the lens galaxy's flux distribution,
and $\mathsf{\Sigma}_{\boldsymbol{d}}$ is the covariance matrix of the data. Some studies choose to subtract the lens galaxy's flux distribution from the imaging data before optimizing a lens model that excludes it \citep[e.g.,][]{bib3:Bolton08}.
Given that the lens flux is loosely related to the lensing phenomenon only through mass-follows-light arguments that are not strictly required, we omit it in the following discussion for brevity.
This likelihood function can be extended in a Bayesian framework to include prior (or regularization) terms on the source \citep{bib3:Warren03, bib3:Treu04, bib3:Koopmans05}, or used to compute the Bayesian evidence \citep{bib3:Suyu06, bib3:Shajib20, bib3:Vernardos22} and perform model comparisons.

\begin{figure*}[!t]
\centering
	\includegraphics[width=1\textwidth]{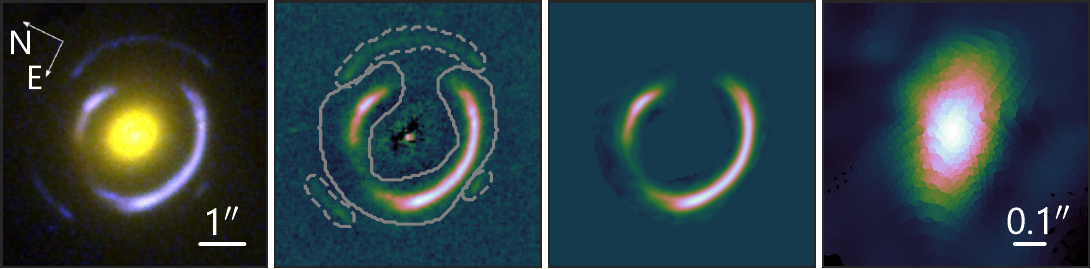}
	\caption{\label{fig3:double_ring}
    \textit{First panel:} False-color image of the system SDSS J0946$+$1006 combining three \textit{HST} filters \citep{bib3:Sonnenfeld12}. \textit{Second panel:} \textit{HST} image in the F814W filter of the system with the galaxy light subtracted. This system has lensed arcs from multiple source galaxies at different redshifts, which are grouped with solid and dashed contours. \textit{Third panel:} Model of the lensed arc from only the brightest source. \textit{Fourth panel:} Corresponding source reconstructed on an adaptive grid using every data pixel and curvature regularization \citep[for the analysis of the data, see Chapter 4 of][and for the modeling method, see \citealt{bib3:Vernardos22}]{bib3:Bayer21}.}
\end{figure*}

When the lensed arcs from an extended source are resolved, each pixel is a constraint for the lens model. 
Given a deflection field $\aang(\tang)$, which depends on the lens mass distribution through $\boldsymbol{\xi}_{\mathrm{mass}}$, the lens equation
\begin{equation} \label{eq3:lens_equation}
    \bang(\tang) = \tang - \aang (\tang) 
\end{equation}
can be used to map any position $\tang$ on the image plane to the corresponding position $\bang$ on the source plane \refedit{\citep[for a detailed explanation of the strong lensing formalism, see][\citealt{bib3:Meneghetti22}, or \chapintroalt]{bib3:Schneider92}}.
We can then easily compute the lensed flux at any location on the image plane as
\begin{equation} \label{eq3:fluxes}
    I(\tang) = S[\bang(\tang)],
\end{equation}
where $S$ is the light distribution of the source that depends on $\boldsymbol{\xi}_{\rm source}$, and we use the fact that lensing conserves surface brightness.
The dependence of $I(\boldsymbol{\theta})$ on the mass through $\bang(\tang)$ is almost always non-linear, which means that we cannot directly \refedit{(i.e., through a linear inversion)} solve \eqref{eq3:lens_equation} to obtain the true parameters $\boldsymbol{\xi}_{\rm mass}$, even when $S$ \refedit{(and equivalently the values of $\xi_{\rm source}$)} is perfectly known.

In practice, $S$ is an unknown that must be solved simultaneously with the lens potential and requires special attention.
The most straightforward choice for it is an analytic function, for example, a S\'ersic profile \citep{bib3:Sersic68}, whose parameters are treated in the same forward-modeling way as for the lens potential. However, several more advanced techniques have been developed over the years that allow a free-form reconstruction of the source, each with its advantages and disadvantages. The semi-linear inversion technique of \citet{bib3:Warren03} uses a regular grid of pixels to approximate the source brightness.
\refedit{
Although the degrees of freedom are much higher in this case, the use of regularization \refedit{or prior} terms of a specific form can greatly facilitate obtaining the best-fit solution.
If a prior is used with quadrature terms of the source pixel brightness values, then the derivative of the posterior probability function\footnote{This is also referred to as the penalty or loss function.} can be obtained analytically given $\boldsymbol{\xi}_{\mathrm{mass}}$.}
As a result, the source reconstruction turns into a linear inversion problem for a given set of non-linear parameters within $\{\boldsymbol{\xi}_{\rm mass}, \boldsymbol{\xi}_{\rm source}\}$, where $\boldsymbol{\xi}_{\rm source}$ now refers only to \refedit{non-linear parameters} of the source that are not solved through the linear inversion.

\begin{figure*}[!t]
\centering
	\includegraphics[width=\textwidth]{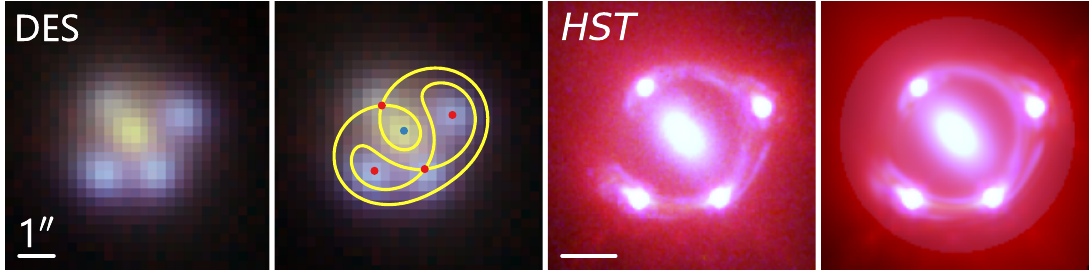}
	\caption{\label{fig3:galaxy_lens_modeling}
	Lens modeling of a galaxy-scale lens system -- here, the lensed quasar WGD J2038$-$4008. \textit{First panel:} The false-color image of the system from the DES $giy$-bands, where only the point images are resolved \citep{bib3:Agnello18}. \textit{Second panel:} Illustration of a lens model based only on the quasar image positions \citep[red points; ][]{bib3:Agnello18}. \refedit{The blue point shows the center of the lensing galaxy, and the yellow contours trace the saddle-point contours on the arrival-time surface.} %
	\textit{Third panel:} \refedit{The false-color image of the system from 3-band \textit{HST} imaging (in F160W, F814W, and F475X filters), where the lensed arcs from the quasar host galaxy can be seen in greater detail.} %
	\textit{Fourth panel:} %
	The \refedit{full-pixel based model} of the \textit{HST} imaging from \citet{bib3:Shajib22}. The white bars represent 1\arcsec.
 }
\end{figure*}

Typical choices of source regularization are gradient and curvature that impose smoothness on the source solution through its derivatives \citep[a standard approach in this kind of problem, e.g.,][]{bib3:Press92}.
\refedit{Alternatively, \citet{bib3:Vegetti14} and \citet{bib3:Nightingale18} use adaptive regularization, which changes the degree of smoothing based on the source brightness, and both studies discuss that this scheme is vital for reconstructing compact sources.
Other} choices include a covariance kernel prior, which is based on observations of the galaxy brightness power spectrum and thus a more physically justifiable choice \citep{bib3:Vernardos22}, or a multi-scale regularization through the use of sparsity constraints on a wavelet representation of the source \citep{bib3:Galan21}.
Choosing a different basis to represent the source can in itself significantly reduce the number of degrees of freedom while still having enough flexibility to represent complex light profiles across different scales \citep[][]{bib3:Birrer15, bib3:Tagore16}. This can be similarly achieved by reconstructing the source on an irregular, adaptive grid that can have increased resolution in the most magnified areas of the source, that is, near the caustics \citep[][]{bib3:Vegetti09, bib3:Nightingale15, bib3:Vernardos22}.
\figref{fig3:double_ring} illustrates an example of lens modeling based on high-resolution \textit{HST} imaging of a lensing system with a bright, extended source component.
The source is reconstructed on an adaptive grid using curvature regularization \citep[Chapter 4 of][]{bib3:Bayer21}.

If there is a point source within the extended source galaxy, then the positions of its point-like images provide additional constraints on the mass model as conjugate points. 
One approach to model the point-like source is to include the location of its multiple images as free parameters and then require the mass model to trace them back to the same location on the source plane through the lens equation.
Alternatively, the point-source location can be free, and the predicted image positions must match the observed ones.
In practice, the choice between the two approaches is based on computational and sub-grid effect arguments \citep[see][for a more detailed presentation on the topic]{bib3:Keeton01}.
\figref{fig3:galaxy_lens_modeling} illustrates an example of lens modeling based on high-resolution \textit{HST} imaging of a lensing system with both a point-like and an extended source component, that is, a quasar and its host galaxy.
In this case, the model for the extended host galaxy assumes an analytic S\'ersic profile and an additional free-form component described by a basis set of shapelets \citep{bib3:Birrer15, bib3:Shajib22}.

\subsubsection{Machine-learning based parameter extraction}
\label{sec3:machine_learning_methods}

Recently, machine learning (ML) based methods have been developed to extract lensing parameters -- for example, the Einstein radius, power-law index, ellipticity, and shear parameters -- from the imaging data \citep[e.g.,][]{bib3:Hezaveh17, bib3:Morningstar19, bib3:Adam22, bib3:Schuldt22}. Some studies also explored reconstructing the lens mass and source flux distributions using machine learning algorithms \citep[e.g.,][]{bib3:Chianese20, bib3:Karchev22, bib3:Mishra22, bib3:Biggio23}. In these approaches, a machine learning algorithm -- usually a neural network -- is trained using synthetic data since real examples of lens systems are not adequate in number for such data-intensive training. This leads to the critical caveat that the mass density profile in the simulated galaxies is based on empirical priors \citep[e.g.,][]{bib3:Hezaveh17} or cosmological simulations \citep[e.g.,][]{bib3:Adam22}. Thus, the inferred parameters are prior-dependent -- either empirical or physical -- in a similar manner that the forward modeling approach depends on the adopted mass model and the associated priors.

Although ML-based inferences have yet to demonstrate their potential in science applications with real data, they are essential for quickly extracting lensing parameters with minimum human supervision. This is crucial for modeling huge samples of lenses like those expected from several large surveys upcoming in the 2020s \citep[e.g.,][]{bib3:Park21, bib3:WagnerCarena21}.
For such large datasets, the traditional forward modeling techniques would be unfeasible due to either computational or human time restrictions. Finally, ML-based inference provides a direct and fast way to constrain quantities of interest that can be otherwise too cumbersome to infer from the data through a traditional approach, for example, detecting individual or populations of dark subhalos or constraining the subhalo-mass function \citep{bib3:Brehmer19, bib3:Diaz20, bib3:Coogan20, bib3:Coogan22, bib3:Ostdiek20, bib3:Ostdiek22, bib3:Vernardos20, bib3:WagnerCarena21, bib3:Zhang22, bib3:Anau22}. Note that this particular science application is discussed in detail in \chapdm.

\subsection{Lens mass models} \label{sec3:lens_mass_models}

Lens mass models can vary in complexity and eventually in the number of free parameters, reflecting that the mass distribution in galaxies is a non-trivial problem for which methodologies and algorithms are still evolving.
Although it is possible to extract lensing information with free-form models and directly connect them to galaxy properties of interest (see \chapintroalt), lens modeling with simple parametrization has been the most common practice in the literature \citep[e.g.,][]{bib3:Ritondale19, bib3:Schmidt23}. We briefly introduce some commonly used models with simple parametrization in \secref{sec3:parametric_lens_models}, discuss the degeneracies that impact them in \secref{sec3:modeling_degeneracies}, and describe free-form models in \secref{sec3:free_form_models}. %

\subsubsection{Simply parametrized mass models} \label{sec3:parametric_lens_models}

\refedit{\citet{bib3:Keeton01b} provides a large catalog of simply parametrized models, which includes the commonly used ones such as the singular isothermal sphere or ellipsoid \citep[SIS or SIE;][]{bib3:Kormann94}, the \refedit{pseudo-isothermal elliptical mass distribution \citep[PIEMD;][]{bib3:Kassiola93}, the softened power-law elliptical mass distribution \citep[SPEMD;][]{bib3:Barkana98}, and the Navarro--Frenk--White \citep[NFW;][]{bib3:Navarro96, bib3:Navarro97} profile. Defining the ellipticity in the potential makes lensing computation very efficient, even for complex radial profiles, since all the lensing quantities can be obtained either from the potential itself (e.g., for the time delay) or through numerical differentiation of the potential \citep[e.g., for the deflection, convergence, or magnification;][]{bib3:Kovner87, bib3:Golse02}. However, moderately elliptical potentials (e.g., with axis ratio $q \lesssim 0.6$) can lead to unphysical shapes in the convergence \citep{bib3:Kassiola93} and also introduce implicit azimuthal variation or ellipticity gradient in the convergence \citep{bib3:Gomer23}. For simply parametrized profiles with ellipticity defined in the convergence, computing deflection angle becomes computationally expensive without an analytical solution due to needing a 2D numerical integration.  Among such elliptical convergence profiles, \citet{bib3:Tessore15} provide an analytical solution for the power-law radial form. Alternative parameterizations of the elliptical NFW convergence profile have also been devised \citep{bib3:Oguri21, bib3:Heyrovsky24}. \citet{bib3:Shajib19b} provides a computationally efficient general solution for any radial form using a superposition of elliptical Gaussian components \citep[i.e., the multi-Gaussian expansion;][]{bib3:Emsellem94, bib3:vandeVen10}.}}

\refedit{On top of such a simply parametrized profile describing the primary lens mass distribution, it} is also often necessary to include a constant \refedit{shear field} (often shortened as XS) lensing potential to accurately model the distortions in the lensed arcs or the Einstein ring. \refedit{These additional distortions can arise from nearby perturbers and large-scale structures \citep{bib3:Keeton97} or the additional angular structure in the central lensing galaxy or galaxies \citep{bib3:Witt96, bib3:Hilbert07, bib3:Gomer18, bib3:Gomer21, bib3:Barrera21, bib3:Vandevyvere22b, bib3:Vandevyvere22a, bib3:Etherington23}. This constant shear field is commonly referred to as `external shear' in the literature, pointing to the former of the two origins mentioned above. However, we recommend using the term `residual shear' instead as a more general terminology. The magnitude of residual shear commonly exceeds 0.1, which is difficult to explain if this shear originates solely from the line-of-sight structures.} Until the mid-2010s, the SIE+XS model has been the most popular choice to model large samples of galaxy-scale lenses \citep[e.g.,][]{bib3:Bolton08b, bib3:Sonnenfeld13}, adequate for the science application requirements at the time.
Although simple SIE models can sufficiently constrain the Einstein radius, obtaining other important properties, such as the radial slope of the mass profile, requires models with additional degrees of freedom.
Improved data quality and analysis techniques have allowed the use of such models \citep[e.g.,][]{bib3:Ritondale19, bib3:Shajib21}.%

Simply parametrized lens models with larger degrees of freedom, for example, a superposition of a stellar component with a constant or varying mass-to-light ratio and a dark matter component usually described by the NFW profile, have been adopted by some studies \citep[e.g.,][]{bib3:Treu10,bib3:Sonnenfeld15, bib3:Oldham18, bib3:Shajib21}. Although necessary for the addressed science questions, such mass profiles with more free parameters amplify the impact of degeneracies inherent to lensing. It is thus often necessary to constrain these additional parameters by incorporating non-strong-lensing observables such as stellar kinematics and weak lensing, or by incorporating informative priors \citep[e.g.,][]{bib3:Sonnenfeld18, bib3:Shajib21}. \refedit{We note that \citet{bib3:Sonnenfeld18} and \citet{bib3:Shajib21} combined strong-lensing information with stellar kinematics or weak lensing using `summary observables', such as the Einstein radius or the reduced slope $\xi_{\rm rad}$ (see \secref{sec3:modeling_degeneracies} for definition), instead of simultaneously fitting the abovementioned complex model to the full lensing information in the imaging data.}

While the additional degrees of freedom beyond the simple SIE model focus on the radial profile of the lens in most cases, azimuthal structures such as disky-ness, boxiness, ellipticity gradients, or isodensity twists may leave noticeable imprints in the lensed images. For point-like sources, the image flux ratios are the most susceptible to perturbations \citep{bib3:Moeller03, bib3:Keeton03, bib3:Keeton05}. In contrast, for extended sources, the imprint is generally more subtle and detectable only from high-resolution imaging data \citep{bib3:Vandevyvere22b, bib3:Vandevyvere22a}. Fortunately, most of those structures arise from baryonic physics and may also be detectable in the luminosity profile of the lens \citep{bib3:Vandevyvere22a}.

\subsubsection{Common degeneracies in simple parametric modeling} \label{sec3:modeling_degeneracies}

The degeneracies in lens modeling -- both intrinsic in the data and stemming from the parametrization scheme -- are discussed in \chapintro. Here, we summarize the common degeneracies that largely impact simply parametrized lens models for the convenience of the readers.

The MSD, which is intrinsic to imaging observables in lensing, originates from the mass-sheet transform \citep[MST,][]{bib3:Falco85, bib3:Saha00}
\begin{equation}
	\kappa \to \kappa^{\prime} = \lambda \kappa + 1 - \lambda,
\end{equation}
where $\lambda$ is a constant.
This equation implies a source position transformation \citep[e.g.,][]{bib3:Schneider14}, where the unknown source position is altered as
\begin{equation}
	\beta \to \beta^{\prime} = \lambda \beta.
\end{equation}
A more general but approximate degeneracy happens when $\lambda$ is not a constant anymore but depends on the position, that is, $\lambda \equiv \lambda(\btheta)$ \citep{bib3:Unruh17, bib3:Wertz18}.

\refedit{Most simply parametrized mass profiles artificially limit the MSD. For example, this is easily demonstrable for the power-law mass model, as the MST of a power law is mathematically not a power law anymore. Although MST-invariant quantities exist that the imaging observables can constrain \citep[e.g.,][]{bib3:Wagner17, bib3:Wagner18}, commonly employed mass models are not parametrized based on those quantities. The standard practice in the literature to assume simply parametrized models, such as the power law, is usually validated on numerous non-lensing constraints demonstrating that the power-law model is a `good' approximation for elliptical galaxy mass profiles \citep[e.g.,][]{bib3:Thomas07, bib3:Tortora14, bib3:Bellstedt18}. The model-independent radial quantities constrained by the imaging data are the Einstein radius} $\tae$ and the MST-invariant \refedit{`reduced slope'} defined as
\begin{equation} \label{eq3:mst_invariant_imaging_observable}
	\xi_{\rm rad} \propto \frac{\tae \alpha^{\prime\prime} ({\tae}) }{1 - \kappa ({\tae})},
\end{equation}
where $\alpha^{\prime\prime}$ is the second derivative of the deflection angle [see Eq. 42 of \citet{bib3:Birrer21b} for the full definition of $\xi_{\rm rad}$, also \citet{bib3:Kochanek21}]. For a power-law convergence profile $\kappa (\theta) \propto \theta^{-\gamma_{\rm PL} + 1}$, this quantity becomes $\xi_{\rm rad} = \gamma_{\rm PL} - 2$. As a result, the choice of a SIE+XS lens model fixes $\xi_{\rm rad} = 0$ and only extracts $\tae$ from the imaging data.

Degeneracies in lens modeling can also arise from particular parametrizations of the lens model. One such example is the shear--ellipticity degeneracy, as the total shear can be redistributed between the `internal' shear, arising from the ellipticity of the central deflector, and the external shear \citep{bib3:Kassiola93}. Specifically, a SIE+XS model can be modeled with only an SIE model that has a quadrupole moment equaling 1/3\textsuperscript{rd} of the shear \citep{bib3:An05}. This degeneracy is particularly apparent when modeling with point-like image positions as the only constraints \citep{bib3:Witt96}.

\subsubsection{Free-form models} \label{sec3:free_form_models}
In the most commonly used lens models, the main focus is on
capturing the radial shape of the mass profile, while any azimuthal structure beyond an elliptical shape with \refedit{residual shear} is of secondary importance.
Real galaxies can have more complicated mass profiles, with higher order moments present, such as disky-ness, boxiness, or bar- or disk-like components \citep[e.g.,][]{bib3:Trotter00,bib3:Claeskens06,bib3:Hsueh17,bib3:Frigo19}, radial dependence of the ellipticity or the orientation of the isodensity contours \citep[i.e., \refedit{ellipticity} gradients, twists, or lopsidedness, e.g.,][]{bib3:Hao06,bib3:Nightingale19, bib3:Barrera21}, or even features that do not fit into a simple parametric description, such as merger products that are not yet completely relaxed and populations of substructures (e.g., satellite subhalos or perturbers along the line of sight).
Detecting such deviations from the simple parametric profiles depends on numerous factors, such as their alignment with the smooth potential \citep{bib3:Vandevyvere22a}, the complexity of the brightness profile of the source \citep{bib3:Vernardos22}, the signal-to-noise ratio, etc.
However, multipole components in the mass potential beyond the combined effect of ellipticity and \refedit{residual shear} have been recently detected in very-long-baseline interferometric observations of a strong lens \citep{bib3:Powell22}.
Although not accounting for such structures can bias the mass model by up to several percent -- which is mostly acceptable except for time-delay cosmography (see \citealt{bib3:Birrer24}) and for detecting dark matter substructure (see \chapdmalt) -- their detection holds valuable information on the formation history and evolution of galaxies.

To this extent, free-form techniques have been developed that either entirely dismiss any parametric mass component and employ a grid of mass pixels to describe the lens potential \citep{bib3:SahaWilliams97}, or retain a parametric model as a first-order smooth component and combine it with a similar pixel grid that now focuses specifically on capturing higher order deviations \refedit{\citep{bib3:Koopmans05, bib3:Suyu10}}.
In both cases, regularization assumptions or priors on the free-form pixel grid are necessary to obtain a solution and to prevent the appearance of unphysical mass distributions.
Existing techniques are based on forward models %
or extensions of the semi-linear inversion technique \citep{bib3:Vernardos22}.

The specific form of the regularization priors plays an important role in the quality of the obtained solutions.
It has been shown that purely mathematically motivated priors (e.g., curvature) can lead to biased potentials as opposed to more physically driven ones, based on the observed light properties of real galaxies \citep[e.g.,][]{bib3:Vernardos22}.
\citet{bib3:Galan22} proposed a wavelet-based regularization technique that finds solutions that satisfy sparsity constraints. 
\citet{bib3:Biggio23} completely replaced the pixel grid with a neural network. %
These approaches allow the use of purely data-driven regularization that is the most compatible with the data (without computing and comparing the Bayesian evidence).
It remains to be seen how well these new and promising techniques can perform on real data and robustly recover deviations from smooth, parametric models that encode galaxy evolution.

\refedit{One way to allow the prior to be less informative for free-form models is to marginalize over an ensemble of solutions. Such ensembles were first introduced for free-form models made up of mass tiles or pixels \citep{bib3:Williams00, bib3:Saha04, bib3:Coles14}. In these and related works, the mass distribution is required to be non-negative and centrally concentrated in a broad sense. Within these prior conditions, models that correctly reproduce the observed positions of point-like image features are randomly sampled to form the ensemble of solutions. This ensemble of solutions is then effectively the posterior of the model parameters, which include all the mass pixel values, and the posteriors of model-predicted quantities can also be obtained from this ensemble \citep[e.g.,][]{bib3:Williams00}. The ensemble can be further filtered according to how well the whole image or the pixels on the lensed arcs from the extended source can be fitted \citep[e.g.,][]{bib3:Denzel21b}.}

\refedit{
Free-form models naturally allow a broad range of mass profile shapes, both radially and azimuthally, thus exploring the degenerate space of the mass profile shapes. The same effect can be obtained with simply parametrized models by combining posteriors from models with different parametric forms, albeit to the limited extent allowed by the variety of the adopted parametrizations  \citep[e.g.,][]{bib3:Suyu14, bib3:Birrer19, bib3:Shajib22}.}

\subsection{Bayesian hierarchical framework}\label{sec3:hierarchical_framework}

The Bayesian hierarchical framework can be used to constrain the population properties from a sample of strong lenses \citep[e.g.,][]{bib3:Sonnenfeld15}. This framework also allows one to incorporate a selection function of the lensing galaxies and generalize the sample properties to the population of all galaxies that are of the same type as the lensing ones \citep[e.g.,][]{bib3:Sonnenfeld19b}. Within the hierarchical analysis, there are two levels of parameters: hyper-parameters that dictate the distribution of the parent population of the lens galaxies and parameters pertaining to individual lens galaxies sampled from the parent population. The hierarchical framework connects \refedit{the population-level hyper-parameters to the observed data through the individual-galaxy-level parameters}. According to the Bayes' theorem, the posterior probability distribution of the hyper-parameters $u$ is given by
\begin{equation}
    p(u \mid \mathcal{D}) \propto p(\mathcal{D} \mid u) \ p(u),
\end{equation}
where $\mathcal{D}$ is the dataset, $p(\mathcal{D} \mid u)$ is the likelihood, and $p(u)$ is the prior. For model parameters $w_i$ pertaining to individual galaxies, the above equation can be expressed as
\begin{equation} \label{eq3:hierarchical_posterior}
    p(u \mid \mathcal{D}) \propto p(u) \prod_i \int \mathrm{d}w_i \ p(\mathcal{D}_i \mid w_i) \ p(w_i \mid u),
\end{equation}
where $\mathcal{D}_i$ is the data from to the $i$-th individual lens galaxy. %
This approach can infer any property at the population level with the associated mean and scatter values. See, for example, \citet{bib3:Sonnenfeld21} for a detailed presentation on the hierarchical framework with specific examples of hyper-parameters $u$.

\subsection{Incorporating non-strong-lensing observables} \label{sec3:non_lensing_observables}

Incorporating non-strong-lensing observables can be used to break the degeneracies in strong lensing analysis. For example, stellar kinematics data can be used to break the MSD \citep[e.g.,][]{bib3:Romanowsky99, bib3:Treu02b,bib3:Treu04,bib3:Shajib18, bib3:Birrer20, bib3:Shajib21, bib3:Tan24}, whereas spectroscopic stellar population analysis \citep[e.g.,][]{bib3:Spiniello11}, weak lensing  \citep[e.g.,][]{bib3:Gavazzi07,bib3:Sonnenfeld18c, bib3:Shajib21}, and microlensing \citep[e.g.,][]{bib3:Schechter14, bib3:Oguri14} information can help mitigate the degeneracy between the stellar and dark matter distributions. \refedit{Here, we briefly describe combining strong lensing with stellar kinematics (\secref{sec3:combine_kinematics}) and weak lensing (\secref{sec3:combine_wl}).}

\subsubsection{Combining stellar kinematics with strong lensing} \label{sec3:combine_kinematics}
Imaging observables probe the 2D mass distribution of the lens projected on the plane of the sky, whereas stellar kinematics probe its full 3D mass distribution. Thus, a combination of the two helps break the MSD to robustly constrain the mass distribution in galaxies. Although elliptical galaxies are triaxial, assuming spherical symmetry has been a standard practice for the case of a single aperture-integrated stellar velocity dispersion measurements \citep[see][for a discussion on the impact of this assumption]{bib3:Sonnenfeld12}. Then, the stellar velocity dispersion is obtained by solving the spherical Jeans equation
\begin{equation} \label{eq3:jeans_equation}
	\frac{{\rm d} \left( l(r)\ \sigma_{\rm r}^2 \right)}{{\rm d} r} + \frac{2 \beta_{\rm ani}(r)\ l(r) \  \sigma_{\rm r}^2}{r} = - l(r)\ \frac{{\rm d} \Phi}{{\rm d} r}.
\end{equation}
Here, $l(r)$ is the 3D luminosity density of the stars, $\sigma_{\rm r}$ is the intrinsic radial velocity dispersion, and $\beta_{\rm ani}(r)$ is the anisotropy parameter relating $\sigma_{\rm r}$ with the tangential velocity dispersion $\sigma_{\rm t}$ given by
\begin{equation} \label{eq3:anisotropy_parameter}
	\beta_{\rm ani}(r) \equiv 1 - \frac{\sigma_{\rm t}^2}{\sigma_{\rm r}^2}.
\end{equation}
By solving the Jeans equation, the line-of-sight velocity dispersion, which is the kinematic observable, is obtained as
\begin{equation} \label{eq3:los_velocity_dispersion}
	\sigma_{\rm los}^2(R) = \frac{2G}{I(R)} \int_R^{\infty} \mathcal{K}_{\beta} \left(\frac{r}{R} \right) \frac{l(r)\ M(r)}{r} \ {\rm d} r
\end{equation}
\citep{bib3:Mamon05}. Here, $M(r)$ is the 3D enclosed mass within radius $r$. The function $\mathcal{K}_{\beta}(u)$ depends on the parameterization of $\beta(r)$ [see \citet{bib3:Mamon05} for specific forms of $\mathcal{K}_{\beta}(u)$ corresponding to different $\beta(r)$]. Thus, the observed velocity dispersion in \eqref{eq3:los_velocity_dispersion} can be written as a function of the lens model parameters as
\begin{equation}
	\sigma_{\rm los}^2 = \frac{\ds}{\dds}\ c^2\ J(\boldsymbol{\xi}_{\rm mass},\ \boldsymbol{\xi}_{\rm light},\  \beta_{\rm ani},\ \lambda),
\end{equation}
where $\boldsymbol{\xi}_{\rm mass}$ are the deflector's mass model parameters, $\boldsymbol{\xi}_{\rm light}$ are the deflector's light model parameters  \citep{bib3:Birrer16}. In this form, the function $J$ is independent of cosmology. It only depends on the lens model parameters $\{\boldsymbol{\xi}_{\rm mass},\ \boldsymbol{\xi}_{\rm light}\}$, the anisotropy profile $ \beta_{\rm ani}(r)$, and the MST parameter $\lambda$. All the cosmological dependence of $\sigma_{\rm los}$ is contained in the distance ratio $\ds/\dds$. To reproduce the observed velocity dispersion integrated within an aperture, the computed luminosity-weighted velocity dispersion needs to be blurred with the PSF $\mathcal{P}$ as
\begin{equation}
	\sigma^2_{\rm ap} = \frac{\int_{\rm ap} \left[ I(R)\ \sigma^2_{\rm los}(R) \right] * \mathcal{P}  \ R \ \mathrm{d}R \mathrm{d}\theta}{ \int_{\rm ap} I(R) * \mathcal{P} \ R \ \mathrm{d}R  \mathrm{d}\theta},
\end{equation}
where the $*$ symbol denotes the convolution operation. To obtain analytic solutions for specific choices of mass, light, and anisotropy profiles, see \citet{bib3:Koopmans06b} for the case of power-law mass and light profiles with constant anisotropy, and \citet{bib3:Suyu10} for the case with power-law mass profile, the Hernquist light profile \citep{bib3:Hernquist90}, and isotropic stellar orbits.

The constraints from the velocity dispersion measurement can be folded in the lens model posterior with a multiplicative likelihood term
\begin{equation}
    \mathcal{L}_{\rm kin} \propto \exp \left[ -\frac{(\sigma_{\rm ap}^{\rm obs} - \sigma_{\rm ap}^{\rm model})^2}{2 \sigma^2_{\sigma_{\rm ap}^{\rm obs}}}\right],
\end{equation}
where $\sigma_{\sigma_{\rm ap}^{\rm obs}}$ is the uncertainty in the observed velocity dispersion $\sigma_{\rm ap}^{\rm obs}$. Whereas the imaging observables from strong lensing cannot constrain the MST parameter $\lambda$, the stellar velocity dispersion constrains $\lambda$ through the likelihood $\mathcal{L}_{\rm kin}$ \citep[see][for a detailed discussion within a hierarchical Bayesian framework]{bib3:Birrer24}.

Although integrated velocity dispersion from long-slit spectra is the most commonly used kinematic observable in strong lensing studies, a few studies have also incorporated spatially resolved velocity dispersion, mainly from integral field unit (IFU) spectra \citep[e.g.,][]{bib3:Barnabe07,bib3:Barnabe11, bib3:Czoske12, bib3:Spiniello15}.

\subsubsection{Combining weak lensing with strong lensing} \label{sec3:combine_wl}

Weak lensing measures the excess shear quantity (${\rm \Delta} \Sigma$) from tidal distortions of galaxies far away ($\gtrsim$10 arcsec) from the central lensing galaxy. Thus, weak lensing provides information on the mass distribution of the lensing galaxy's outer region, that is, where the dark matter halo dominates. As a result, the weak lensing information constrains the normalization of a dark matter component in the lens mass model. strong-lensing observables provide information on the total mass distribution (e.g., the enclosed mass). Still, there remains a degeneracy between the luminous and dark matter fractions within the total enclosed mass. If a specific profile is assumed for the dark matter distribution, weak lensing data can break this degeneracy between luminous and dark component normalizations \citep{bib3:Sonnenfeld18, bib3:Shajib21}.

The weak lensing information is \refedit{often convenient or appropriate} to be incorporated within the hierarchical framework, \refedit{although there are examples of such combination without using a hierarchical framework \citep[e.g.,][]{bib3:Gavazzi07}}. Since the weak lensing signal from one single lensing galaxy does not usually have enough constraining power on the dark matter normalization, it is often required to stack weak lensing signals from a large sample of elliptical galaxies, \refedit{which are not-necessarily lensing galaxies}. If it is justified to assume that this large sample of elliptical galaxies and the lens galaxy sample under consideration are subsamples of the same parent population, then the weak lensing observables $\mathcal{D}_{\rm weak}$ and the strong-lensing observables $\mathcal{D}_{\rm strong}$ can be jointly considered in the posterior probability function of the population hyper-parameters $u$ as
\begin{equation}
	p(u \mid \mathcal{D}_{\rm strong}, \mathcal{D}_{\rm weak} ) \propto p(\mathcal{D}_{\rm strong} \mid u) \ p(\mathcal{D}_{\rm weak} \mid u) \ p(u).
\end{equation}
Here, the strong lensing likelihood $p(\mathcal{D}_{\rm strong} \mid u)$ can then be expanded using the single system likelihoods similar to the product term in \eqref{eq3:hierarchical_posterior}. Since from Bayes' theorem, we have
\begin{equation}
	p(\mathcal{D}_{\rm weak} \mid u)\ p(u) \propto p(u \mid \mathcal{D}_{\rm weak}),
\end{equation}
it is also valid for numerical convenience to first obtain the posterior of $u$ from weak lensing observables only and then fold this posterior as the prior of $u$ in the hierarchical analysis with only strong-lensing data.

\section{Applications in galaxy properties and evolution} \label{sec3:science_applications}

This section describes what we can learn about galaxy structure and evolution using the lensing galaxy properties. Since strong-lensing galaxies are typically massive ellipticals, most of the strong-lensing studies in the field relate to this type of galaxy (\secref{sec3:galaxy_density_profile}--\secref{sec3:central_images}). However, we briefly discuss strong lensing by spiral galaxies at the end of this section in \secref{sec3:spiral_galaxies}.

\subsection{Galaxy mass density profile} \label{sec3:galaxy_density_profile}

All galaxies are believed to form and grow inside their dark matter halos. Thus, a massive galaxy's total mass density profile comprises two components: the baryonic matter distribution, which includes stars and gas, and the dark matter distribution.  As seen from the comparison of the observed galaxy luminosity function and the distribution of simulated dark matter halos \citep[i.e., abundance matching, see][]{bib3:Moster10}, the stellar-mass fraction decreases with mass, for the mass range of typical lensing galaxies. This is attributed to AGN feedback and is evident from simple comparisons of the stellar and total mass in lensing galaxies \citep{bib3:Auger09, bib3:Kung18}.

\begin{figure*}
	\includegraphics[width=\textwidth]{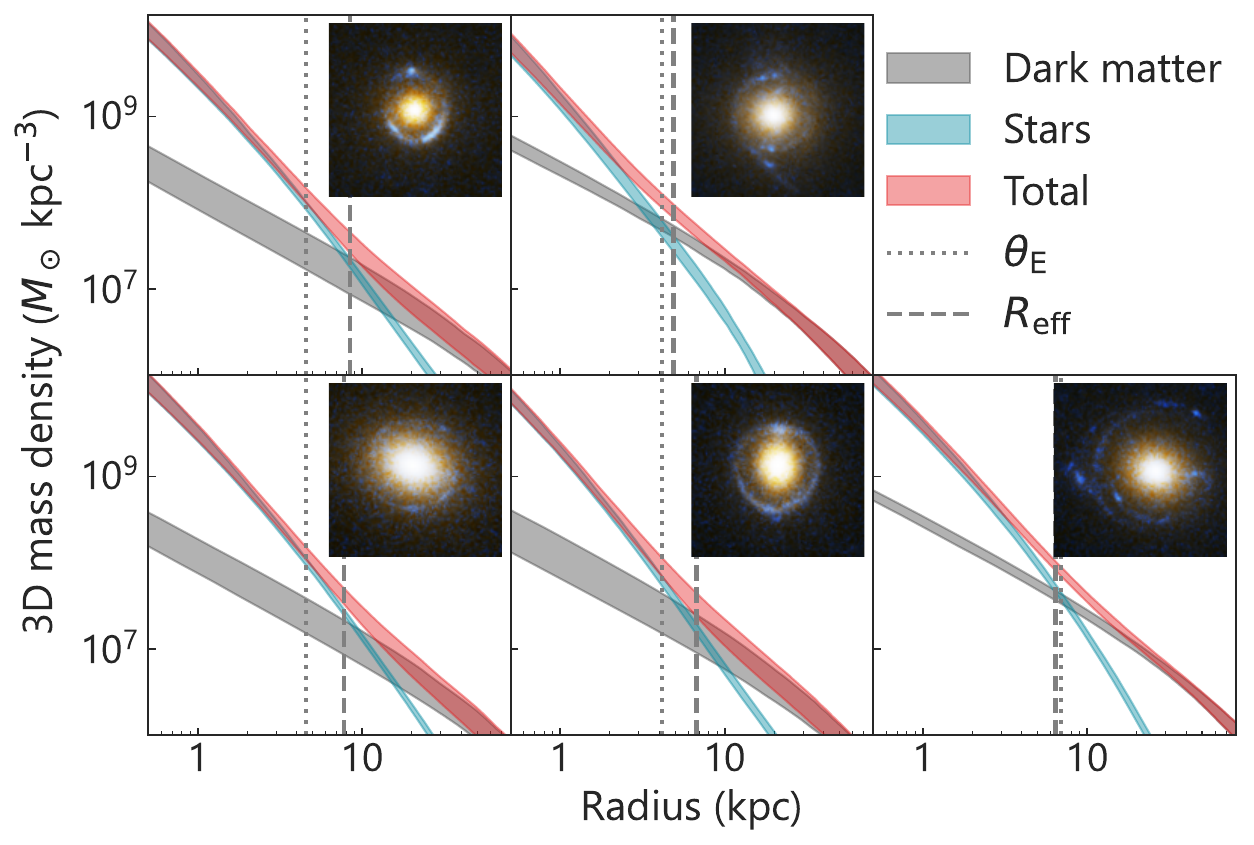}
	\caption{\label{fig3:3D_density_profile}
	Examples of 3D mass distribution in dark matter and stars (i.e., baryons) in lensing elliptical galaxies \citep{bib3:Shajib21}. The false-color image created from the \textit{HST} imaging of each system is illustrated in the inset (\textit{image credit: NASA/ESA, A.~Bolton, and the SLACS team}). \citet{bib3:Shajib21} combined this \textit {HST} imaging data with the stellar kinematics and weak lensing information to decouple the stellar (teal) and dark matter (grey) components of the total mass density profile (red). Only five examples are illustrated here out of the 23 analyzed lenses from the SLACS sample. The vertical dotted and dashed lines mark the Einstein radius $\tae$ and the half-light or effective radius $R_{\rm eff}$, respectively. In most cases, the total density profile is close to the power-law profile, with occasional deviations appearing near the center or far outside the Einstein radius.}
\end{figure*}

\refedit{A power-law model $\rho(r) \propto r^{-\gammapl}$ close to the isothermal case (i.e., $\gammapl \sim 2$) has been found to be sufficient to describe several lensing and non-lensing observables to the noise level}; for example, from strong lensing only or in combination with stellar dynamics \citep{bib3:Kochanek95, bib3:Treu04, bib3:Dye05, bib3:Koopmans06, bib3:Koopmans09, bib3:Barnabe09, bib3:Auger10b, bib3:Dutton14, bib3:Ritondale19, bib3:Powell22, bib3:Tan24}, from the combination of strong and weak lensing \citep{bib3:Gavazzi07}, from stellar dynamics only \citep{bib3:Bertin93, bib3:Gerhard01, bib3:Thomas07, bib3:Tortora14, bib3:Cappellari15, bib3:Bellstedt18, bib3:Derkenne21}, and from X-ray luminosity \citep{bib3:Humphrey10}. This phenomenon -- that the total mass profile in ellipticals approximately follows the power law, whereas neither the baryonic nor the dark matter component individually follows the power law  -- is referred to as the `bulge--halo conspiracy' \citep{bib3:Treu04}, similar to the `disk--halo conspiracy' in spiral galaxies \citep{bib3:vanAlbada86}. \figref{fig3:3D_density_profile} illustrates the dark and luminous components constrained by combining kinematic and weak lensing information with strong lensing for five SLACS lens galaxies \citep{bib3:Shajib21}. In most cases, the total density profile is very close to a power-law form, with deviations only being prominent far from the half-light radius. Galaxy formation simulations suggest that the close-to-isothermal nature of the total density profile originates from rearranging the mass distribution through collisionless accretion in gas-poor mergers \citep{bib3:Johansson09, bib3:Remus13}. %

Analysis of the SLACS lenses finds the mean logarithmic slope $\langle\gammapl\rangle = 2.08 \pm 0.03$ with an intrinsic scatter of $0.16\pm0.02$ from a sample of 85 galaxy--galaxy lenses \citep{bib3:Koopmans09, bib3:Auger10b}. The median redshift of the SLACS lenses is  $\langle z_{\rm SLACS} \rangle \simeq 0.19$. The SIE lens model was adopted in this analysis to constrain $\tae$ from the imaging data. Then, the power-law index $\gammapl$ was obtained from the stellar velocity dispersion measured by the SDSS. \citet{bib3:Shajib21} re-analyzed 23 systems from the SLACS sample with a power-law model instead of the SIE model to obtain the mean logarithmic slope $\langle \gammapl \rangle = 2.08 \pm 0.03$ with an intrinsic scatter of $0.13 \pm 0.02$. \citet{bib3:Ritondale19} analyzed 17 galaxy--galaxy lens systems from the \refedit{BELLS GALLERY} sample with a power-law mass model to find the average logarithmic slope $\langle \gammapl \rangle = 2.00 \pm 0.01$. The mean redshift of the \refedit{BELLS GALLERY} sample is approximately \refedit{$\langle z_{\rm BG} \rangle \simeq 0.5$}.

\begin{figure*}
\centering
\includegraphics[width=0.7\textwidth]{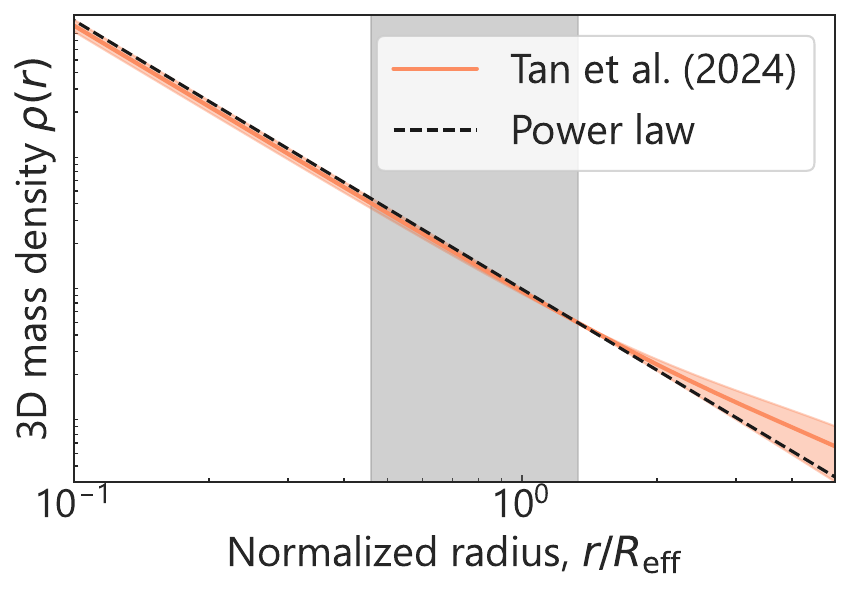}
	\caption{\label{fig3:dinos-i}
	\refedit{The shape of the mean 3D total mass profile for the lens galaxies from the SLACS and SL2S samples constrained by \citet[][\textit{cf.} Fig. 11 therein]{bib3:Tan24} from a joint lensing--dynamics analysis. The orange line shows the population mean with the shaded region representing the 68\% (1$\sigma$) credible region. The dashed black line traces the fiducial power-law mass distribution. The vertical grey shaded region shows the 1$\sigma$ range of the Einstein radius distribution of the sample. For the dynamical modeling, these authors adopted a spatially constant stellar anisotropy profile, consistent with the spatially resolved velocity dispersion measurements of local elliptical galaxies (see references therein). The 3D mass density profile $\rho(r)$ along the vertical axis is normalized at a reference radius and thus has no units in this illustration. The population mean of the mass density profile is consistent with the power-law model within 1$\sigma$.}
	}
\end{figure*}

\refedit{Project Dinos \citep{bib3:Tan24} reanalyzed multi-band \textit{HST} imaging for a sample of $\sim$50 lenses from the SLACS and SL2S samples and then combined the lensing constraints with the stellar kinematics to directly constrain any potential deviation from the power-law profile, where the deviation is parametrized with the internal mass-sheet transformation parameter $\lambda_{\rm int}$ after correcting for the line-of-sight effects (i.e., the external convergence). The `internal' mass-sheet transformation is referred to as such to differentiate it from the effect of the external convergence that acts as an `external' mass sheet. These authors find the power-law model is consistent with both the lensing and kinematic observables within $1\sigma$ for the baseline choice of spatially constant anisotropy profile that is informed by exquisite IFU kinematics of local elliptical galaxies (\figref{fig3:dinos-i}).}

The surface mass in elliptical galaxies constrained from strong lensing can explain the origin of the so-called tilt of the fundamental plane, that is, the tight correlation between the effective radius, the effective surface brightness, and the central velocity dispersion \citep{bib3:Ciotti96}. If the surface mass is used instead of the surface brightness, then the resulting mass plane (in place of the fundamental plane) is not tilted \citep{bib3:Bolton08b}. This result implies that the tilt of the fundamental plane stems from the increase in the dark matter fraction with increasing velocity dispersion or dynamical mass \citep{bib3:Auger09}.

In summary, the total mass profile in galaxies seems to be well described by a power law. \refedit{However, there is also a recent indication of a potential departure from the power law suggested by the non-correlation between the logarithmic slopes from lensing-only and lensing--dynamics analyses \citep{bib3:Etherington23b}.}  The implications of this finding for the individual baryonic and dark matter components are discussed separately below.

\subsubsection{Luminous (baryonic) mass profile} \label{sec3:luminous_structure}

In recent years, spatially resolved spectroscopic surveys of nearby ellipticals using IFUs and high-resolution \textit{HST} imaging have considerably advanced our understanding of the structure and evolution of these galaxies \citep{bib3:Cappellari16}. Such detailed studies of intermediate redshift galaxies ($z\sim0.2-0.7$) using direct observations are impossible. However, the mass of galaxies acting as strong lenses at these redshifts can be mapped out in detail, thereby providing critical information about the progenitors of present-day galaxies and their evolution.

In the last decade, imaging and kinematic data have led to a revision of galaxy classification. The modern analyses classify ellipticals into fast and slow rotators, and those with and without central cores, resulting in four classes \citep{bib3:Cappellari16, bib3:Krajnovic20}. Massive ellipticals ($M\gtrsim 10^{11}\msol$) in the local Universe tend to have small cores with size 0.02--0.50 kpc \citep{bib3:Krajnovic20}, that is, much smaller than the Einstein radius of typical lenses. In contrast, less massive ones appear to have cuspy central light profiles. Cores are believed to result from in-spiraling super-massive black hole (SMBH) binaries, which transfer their angular momentum outward to stars and leave a flatter density core at the center. The population of massive elliptical galaxies is also more likely to be composed of slow rotators, being also morphologically rounder than the lower mass fast rotator counter-part \citep[e.g.,][]{bib3:Weijmans14, bib3:vandeSande17}.

Although it is well known that the baryons and the dark matter do not follow the same radial profile, it is unknown \textit{a priori} whether they have the same angular structure or not. Strong lensing allows us to compare the azimuthal distributions of the light and the total mass distributions, thus detecting any possible difference between the angular structures of the dark matter and the baryons. Several studies report a strong correlation in the ellipticity between the matter and light distributions \citep[e.g.,][]{bib3:Koopmans06b, bib3:Gavazzi12, bib3:Sluse12, bib3:Kostrzewa14}. In contrast, some other studies only report weak or no correlation \citep[e.g.,][]{bib3:Keeton98b, bib3:Ferreras08, bib3:Rusu16, bib3:Shajib19, bib3:Shajib21}. Some differences can be attributed to data quality, modeling procedure, or selection effects \citep{bib3:Shajib21}. The major axes are usually well aligned (with position angle difference $\lesssim$10$^\circ$) between the mass and light distributions \citep[e.g.,][]{bib3:Keeton98b, bib3:Kochanek02b, bib3:Koopmans06b, bib3:Treu09, bib3:Sluse12, bib3:Bruderer16, bib3:Shajib19, bib3:Shajib21}. The cases where the major axes of the mass and light do not align within $\sim$10$^\circ$ also generally have large \refedit{residual} shear. These findings suggest that stellar orbits highly misaligned with the potential can only be sustained in non-isolated galaxies (indicated by the large \refedit{residual shear if interpreted as originating from the presence of nearby galaxies}), which is consistent with what is found in simulations \citep{bib3:Heiligman79, bib3:Martinet88, bib3:Adams07, bib3:Debattista15}.
Although the stellar IMF pertains to the luminous structure, we present the strong lensing results on the stellar IMF separately in \secref{sec3:imf_results}.

\subsubsection{Dark matter profile} \label{sec3:dark_structure}

Numerical simulations show that the dark matter halos and the baryonic mass within them initially follow the NFW profile before star formation begins. However, the baryonic gas has to cool down and fall inward for star formation to begin. The contraction in the baryonic matter deepens the gravitational potential. Thus, the dark matter distribution also contracts in response, \refedit{for which the adiabatic contraction scenario can work reasonably well \citep{bib3:Blumenthal86, bib3:Cautun20}}. In the process of adiabatic contraction {of a spherical mass distribution}, the initial radius $r_{\rm i}$ of a dark matter particle and its final radius $r_{\rm f}$ is related as
\begin{equation} \label{eq3:contraction_1}
	r_{\rm i} \ M_{\rm i} (r_{\rm i}) = r_{\rm f} \ M_{\rm f} (r_{\rm f}),
\end{equation}
where $M(r)$ is the enclosed 3D mass within radius $r$ \citep{bib3:Blumenthal86}. However, numerical simulations find that dark matter does not fully respond to the baryonic infall according to the theoretical model of \citet{bib3:Blumenthal86} \citep[e.g.,][]{bib3:Gnedin04, bib3:Abadi10}. \citet{bib3:Dutton07} prescribe a formalism defining a halo response parameter $\nu$ to adjust the degree of contraction (i.e., the response to the baryonic infall) as
\begin{equation}
	r_{\rm i} \equiv \Gamma^{-\nu} (r_{\rm f})\ r_{\rm f},
\end{equation} 
where $\Gamma (r_{\rm f}) \equiv r_{\rm f} / r_{\rm i}$ is the contraction factor. In this formalism, $\nu = 0$ corresponds to no contraction, and $\nu = 1$ corresponds to fully responsive contraction according to the model of \citet{bib3:Blumenthal86}. The simulations of \citet{bib3:Gnedin04} point to $\nu \sim 0.8$, and those of \citet{bib3:Abadi10} to $\nu \sim 0.4$ \citep{bib3:Dutton14}.

\begin{figure}
\centering
	\includegraphics[width=\textwidth]{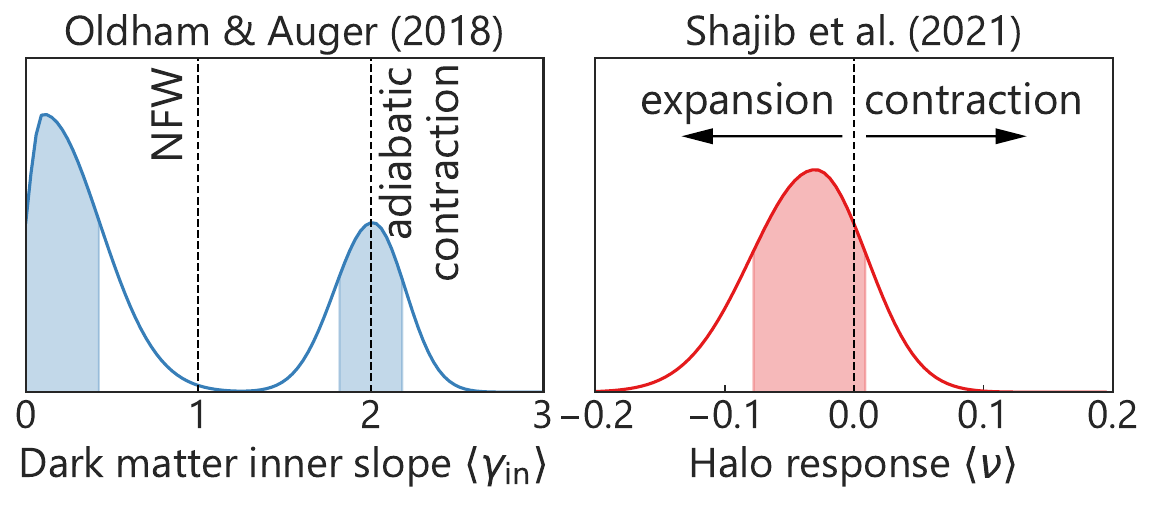}
	\caption{\label{fig3:dm_contraction}
    \refedit{Results on the dark matter distribution from joint lensing--dynamics analyses. \textit{Left-hand panel:} the probability density function for the sample mean of the dark matter inner logarithmic slope $\langle \gamma_{\rm in} \rangle$. This result was obtained from a hierarchical analysis performed on the lensing--dynamics data for 12 strong lenses. The probability distribution shows a bimodality, where one mode is consistent with the adiabatic contraction scenario (i.e., $\gamma_{\rm in} \approx 2$) and the other mode is consistent with an expanded halo (i.e., $\gamma_{\rm in} \approx 0$). \textit{Right-hand panel:} the probability density function of the sample mean of the halo response parameter $\langle \nu \rangle$ obtained from a SLACS subsample of 23 lenses \citep{bib3:Shajib21}. This result is consistent with no contraction or expansion from the regular NFW profile (i.e., $\nu =0$ marked by the vertical dashed line). Although these two results disagree, potential systematics stemming from different parametrizations of the adiabatic contraction or the stellar anisotropy profile are yet to be ruled out as the source of this discrepancy.
    }
    }
\end{figure}

\refedit{\citet{bib3:Dye05} constrained the dark matter profile based on lensing analysis of one lens system and found the inner slope of the dark matter halo to be consistent with the NFW profile.} Several studies have combined additional information, for example, stellar kinematics and weak lensing, with the strong-lensing data to decompose the dark matter and baryonic components from the total density profile constrained by strong lensing. \citet{bib3:Dutton14} adopted multiple dark matter contraction models with fixed $\nu$ to values between $-0.5$ and 1. These authors find that $\nu = 0$, that is, no contraction, best matches the data from the SLACS sample. \citet{bib3:Shajib21} allowed for a fully variable $\nu$ parameter within $-0.5$ and 1 in their dark matter model. These authors find the average contraction in their sample to be $\langle \nu \rangle = -0.03^{+0.04}_{-0.05}$, which is consistent with no contraction and rules out the contraction results from simulations with high statistical significance (\figref{fig3:dm_contraction} right-hand panel). \refedit{In contrast, \citet{bib3:Oldham18} find a cuspier inner logarithmic slope than the NFW profile for the majority of their sample of 12 lens systems, with a smaller subset having shallower ones, pointing to the impact of the environment in the evolution of these galaxies (\figref{fig3:dm_contraction} left-hand panel). However, the potential systematic dependence for all of the above results on modeling choices -- for example, dark matter profile and anisotropy profile parameterizations -- is yet to be investigated thoroughly.}

\subsection{Stellar initial mass function} \label{sec3:imf_results}

When combined with ancillary data, strong lensing provides a method to infer the stellar mass-to-light ratio \citep[$M_{\star}/L$;][]{bib3:Treu10}, which can then be related to the low-mass end of the
stellar initial mass function (IMF) slope in the lensing galaxies \citep[for a review, see ][]{bib3:Smith20}. This is mainly because low-mass stars ($M_{\star}<0.5M_{\odot}$) contribute only by a few percent to the integrated light in the optical, but they give a much larger contribution to the mass \citep{bib3:Conroy12}.

As described in \secref{sec3:galaxy_density_profile}, the total mass distribution in the lensing galaxy can be decomposed into the stellar and dark components by combining lensing and dynamical observables. Thus, the total stellar mass $M_{\star}^{\rm LD}$ can be obtained. An independent method to get the stellar mass $M_{\star}^{\rm SPS}$ of the galaxy is the stellar population synthesis (SPS) method applied on the lens galaxy's photometric or spectroscopic data \citep{bib3:Spiniello11, bib3:Spiniello12}. The $M_{\star}^{\rm SPS}/L$ computed via SPS analysis
depends on the choice of the IMF slope. In particular, a bottom-heavier IMF implies a larger $M_{\star}^{\rm SPS}/L$ because dwarf stars contribute more to the mass than light. Hence, the mismatch parameter:
$\alpha_{\rm IMF} \equiv M_{\star}^{\rm LD} / M_{\star}^{\rm SPS}$.
can be used to infer the lightness or heaviness of the IMF \citep{bib3:Treu10}. For example, if a light IMF -- such as the Chabrier IMF \citep{bib3:Chabrier03} -- is adopted in the SPS method, then $\alpha_{\rm IMF} \sim 1$ would point to the Chabrier IMF being consistent with the lensing and dynamical observables. However, $\alpha_{\rm IMF} \sim 2$ would indicate that the IMF in the lens galaxies is bottom-heavier (i.e., characterized by a larger number of dwarf stars), with a slope more similar to that of a Salpeter IMF \citep{bib3:Salpeter55}, or even steeper.
Note that 
$M_{\star}^{\rm LD}$ 
may depend on lens modeling assumptions, such as the choice of dark matter density profile, the stellar mass-to-light ratio being spatially constant or varying, and the assumed anisotropy profile of the stellar orbits in the dynamical modeling \citep[e.g., see the discussions in][]{bib3:Sonnenfeld18c, bib3:Sonnenfeld19b}. 

\refedit{Although the IMF slope and the low-mass cutoff are degenerate with respect to the lensing data, both can be constrained when combined with dynamics and stellar population analysis.}
\citet{bib3:Barnabe13} show this by studying two strong lenses from the X-Shooter Lens Survey \citep{bib3:Spiniello11} for which both \textit{HST} imaging (for precise lens modeling) and X-Shooter spectra (for stellar population analysis) are available.
Chromatic, microlensing-induced flux anomalies in a galaxy--quasar strong lens can also be used to constrain the stellar IMF \citep{bib3:Schechter14}. This technique is described in \chapmicro, which discusses the theory and applications of microlensing.

\begin{figure}
	\centering
	\includegraphics[width=\textwidth]{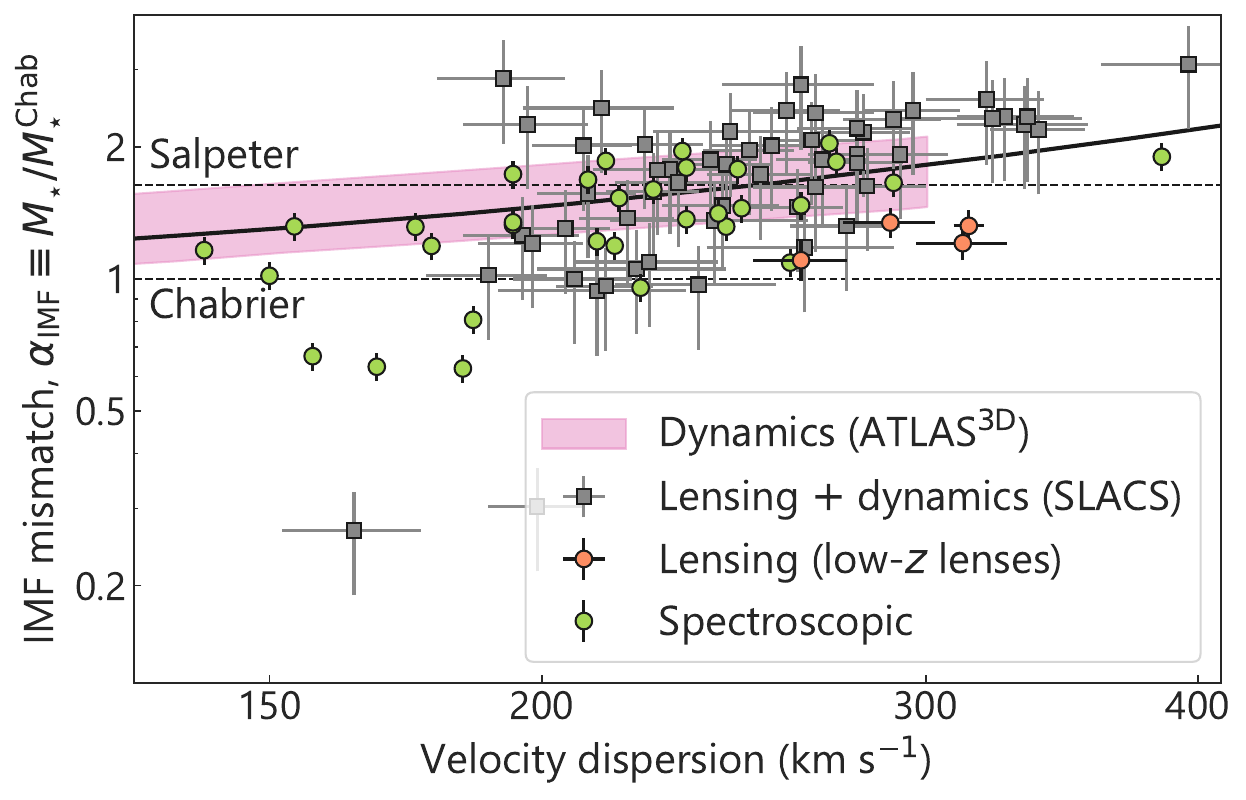}
	\caption{\label{fig3:imf_summary}
	\refedit{Measurements of the stellar IMF from various probes. The pink shaded region traces the 1$\sigma$ intrinsic scatter of the fitted relation from dynamical constraints of the ATLAS$^{\rm 3D}$ sample \citep{bib3:Cappellari13, bib3:McDermid14}. The grey points are based on joint lensing--dynamics analysis for a subsample of the SLACS lenses \citep{bib3:Posacki15}. The orange points are lensing-only measurements from a sample of low-redshift ($z=0.031$--$0.066$) lenses; 3 from the SINFONI Nearby Elliptical Lens Locator Survey (SNELLS) and one discovered from a search on the publicly available MUSE data \citep{bib3:Smith15, bib3:Collier18}. The green points show the fully spectroscopy-based measurements for a sample of 34 elliptical galaxies from \citet{bib3:Conroy12}. The two horizontal dashed lines mark the values expected for Salpeter and Chabrier IMFs. The solid black line illustrates the dependency of the IMF on the velocity dispersion, fitted by \citet{bib3:Posacki15} using the ATLAS$^{\rm 3D}$ and the SLACS samples.}
	}
\end{figure}

Whereas the IMF within the Milky Way is light -- that is, consistent with the Chabrier IMF regardless of the stellar population age and environment \citep{bib3:Chabrier03, bib3:Bastian10} -- the majority of strong lensing studies on elliptical galaxies report consistency with a heavier IMF \citep[e.g.,][]{bib3:Spiniello11, bib3:Sonnenfeld12, bib3:Oldham18}. \refedit{We note, however, that the stellar IMF is degenerate with the choice of the dark matter density profile in most of these studies \citep{bib3:Auger10}.} 
For instance, the SLACS analysis -- by combining lensing and dynamics -- finds the IMF in elliptical galaxies at mean redshift $\langle z \rangle \sim 0.2$ to be consistent with the Salpeter IMF, that is, $\alpha_{\rm IMF} \sim 2$ \citep[see \figref{fig3:imf_summary}; ][also, \citealt{bib3:Grillo09}]{bib3:Treu10}. This result is reproduced with more flexible models for the same SLACS systems or a subset of them \citep{bib3:Auger10b, bib3:Shajib21}. 
This is also in agreement with the more general findings, based on dynamics or SPS analysis only, that the IMF is bottom-heavier for more massive galaxies in general \citep[\figref{fig3:imf_summary};][]{bib3:Cappellari12, bib3:LaBarbera13, bib3:Spiniello14}.%
The general picture is that the low-mass end of the IMF might not be universal across all galaxies, as generally assumed in the last thirty years. However, a consensus on the physical mechanisms responsible for its variation has not yet been reached. According to theoretical work (e.g., \citealt{bib3:Hopkins13, bib3:Chabrier14}), high density, temperature, and turbulence of the gas are key parameters that drive the fragmentation of molecular clouds. Higher density and temperature make the fragmentation easier, forming more dwarf stars, that is, a bottom-heavier IMF.

In contrast with the studies mentioned above on massive, local elliptical galaxies, \citet{bib3:Sonnenfeld19b} find $\log_{10} \alpha_{\rm IMF} = -0.04 \pm 0.11$ with respect to the Chabrier IMF in the SPS method for a sample of strong lensing galaxies at $z \sim 0.6$, which is in tension with \citet{bib3:Shetty14} that reports consistency with a Salpeter IMF for a galaxy at $z\sim1$ from a dynamical analysis. However, allowing spatial gradients in the $M_{\star}/L$ %
can alleviate this tension \citep{bib3:Sonnenfeld18c, bib3:Sonnenfeld19b}. Indeed, a radial gradient in the $M_{\star}/L$, or equivalently a radially varying IMF, has been reported by studies based on the SPS method applied on local massive ellipticals \citep{bib3:Martin-Navarro15, bib3:vanDokkum17,bib3:Sarzi18,  bib3:Barbosa21a, bib3:Barbosa21}. 
These authors find that the central region within $\sim$2 kpc has a heavy IMF (even super-Salpeter), and the IMF in the outer regions gradually becomes light (i.e., Milky-Way-like). 
The current belief is that the IMF is bottom-heavy for stars formed very early in cosmic time via a quick and violent SF burst. These stars usually form a `red nugget' \refedit{\citep[$z\sim2$,][]{bib3:Damjanov11, bib3:Oldham17}}: an ultra-compact red-and-dead massive core. Then, with a second and more time-extended phase, red nuggets merge, interact with other structures in the Universe, and accrete gas. This process causes a growth in size up to a factor of $\sim$5, and only slightly in mass, transforming them into giant local, massive ellipticals. Depending on the merger history of each single galaxy, the red nugget can remain almost untouched in the innermost region, dominating the light there. In this case, this region would have a bottom-heavy IMF. This seems to be the case for NGC~3311, the central galaxy in the Hydra cluster \citep{bib3:Barbosa21a}, M87 \citep{bib3:Sarzi18}, and many other very massive low-$z$ early-type galaxies. However, the red nugget can also be destroyed or contaminated by accreted or lately formed stars. In that case, one would measure a Milky-Way-like IMF, which is the characteristic for stars formed later on and through more time-extended star formation channels. This `two-phase formation scenario' \citep{bib3:Naab09} is also supported by the discovery of `relic galaxies' \citep{bib3:Trujillo14, bib3:Spiniello21}, the local counterparts of red nuggets that somehow wholly missed the size-growth and evolved passively and undisturbed across cosmic time. The IMF for these peculiar and rare objects has been measured to be steep everywhere up to at least one effective radius \citep{bib3:FerreMateu17}. Finally, simulations have lately shown that not all elliptical galaxies formed via this two-phase formation scenario \citep[e.g.,][]{bib3:Pulsoni21}, although this becomes more and more common with increasing stellar mass.

The finding of the Salpeter IMF from a combination of lensing and dynamics in the near-by Universe is consistent with this scenario, as strong-lensing information is sensitive to the galaxy's inner region (typically $\lesssim$6 kpc). However, at higher redshift, lensing probes larger and larger regions, which explains the results presented in \citet{bib3:Sonnenfeld19b}. A radially decreasing $M_{\star}/L$ also explains or alleviates the reported tension between lensing-based studies themselves \citep{bib3:Sonnenfeld18c, bib3:Sonnenfeld19b, bib3:Shajib21}. 

Furthermore, studies of few local massive lenses for which the Einstein radius is much smaller than the effective radius, and hence where the stellar mass dominates the lensing inference, indicate that bottom-heavy IMFs are excluded by lensing \citep{bib3:Ferreras10, bib3:Smith15, bib3:Leier16}. 
\refedit{Allowing for a variable cut-off on the low mass end of the IMF can reconcile the $M_{\star}/L$ measurement from strong lensing of \citet{bib3:Smith15} with the IMF-sensitive absorption line measurements of \citet{bib3:Conroy12}. Still, the discrepancy with other lensing--dynamics measurements remains.}

In conclusion, the currently preferred scenario sees the majority of massive galaxies having a bottom-heavy IMF in their innermost region, where the pristine stellar population dominates that formed at $z>2$ through a star formation burst, while stars in outskirt are distributed by a Milky-Way like IMF. However, depending on the single galaxy's detailed merger tree and cosmic evolution, the IMF can differ from system to system.

In the future, IFU-based stellar kinematics and population analysis of strong lensing galaxies in combination with lensing constraints can shed light on the presence or absence of IMF variations and spatial gradients in ellipticals at intermediate redshifts. However, to properly track any evolution in the IMF properties of elliptical galaxies across redshift, it would be essential to mitigate systematic impacts through a uniform choice of models and to account for selection differences between samples. 

\subsection{Constraints on the very central densities and \refedit{SMBH mass} from central images} \label{sec3:central_images}

Gravitational lensing theory predicts that the number of multiple images must always be odd. In systems with three (or five) images, two (or four) are formed roughly at the Einstein radius from the lens center, which is $\lesssim$2$\arcsec$ for galaxies. The odd 3rd (or 5th) image is formed very near the center of the lens. It is always demagnified, usually significantly, and superimposed on the light of the lensing galaxy, making it hard to detect. \refedit{Since the demagnification of the central image depends on the central density profile, the detections (or the lack thereof) of the central image can constrain or put an upper limit (or lower limit) on the steepness of the inner density profile or the SMBH mass \citep[e.g.,][]{bib3:Winn04}, with the degeneracy between the two broken with stellar light distribution informing the stellar mass profile \citep[e.g.,][]{bib3:Wong15, bib3:Tamura15}.} 

Most existing searches for central images of strongly lensed quasars (given that they are much brighter sources) rely on optical, or radio wavelengths. The radio wavelengths are most favorable as the lensing galaxy is generally transparent in that range, but these investigations are limited by the fact that quasars are usually radio-quiet. Out of $\sim$200 doubles discovered to date, only two have observed central images where the lens is a single galaxy\footnote{There are at least five cases of central images where the lens has two or more main lensing galaxies.}: PMN J1632$-$0033, with the central image demagnified by a factor of 0.004 or 6 magnitudes compared to the brightest image \citep{bib3:Winn04}, and PKS 1830$-$211, with the central image demagnified by a factor of 0.007 or 5.4 magnitudes \citep{bib3:Muller20}. No reliable detection exists of the $\sim$50 known quads with a single lensing galaxy. Upper limits have also been placed. For example, \cite{bib3:Quinn16} place an upper limit of $\sim 10^{-4}$ on the magnification of the central image with respect to the brightest visible image in the double B1030$+$074. 

Upper limits on the flux of the central image have also been placed at X-ray wavelengths for several quads and doubles. The stacking of X-ray monitoring data allows effective exposures of several hundreds of kilo-seconds without any contamination from the lens. However, the data are still too shallow to strongly constrain the lens galaxy's inner density profile. One of the deepest upper limits has been achieved for HE~0435$-$1123 \citep[e.g.,][]{bib3:Chen12, bib3:Guerras2017}.

Detection of central images, or the lack thereof, has been used to place constraints on the central mass density of the lensing galaxies and the mass of the central SMBH \citep{bib3:Mao01, bib3:Rusin01, bib3:Wong15, bib3:Tamura15, bib3:Quinn16, bib3:Perera23}. \refedit{Even without a central image detected, a lensed image that is sufficiently close to the center can be used to measure the SMBH mass. For example, such an SMBH mass ($M_{\rm SMBH} = 3.27\pm2.12 \times 10^{10}$ M$_{\odot}$; 3$\sigma$ confidence limit) was measured using the image at $\sim$1 kpc distance from the center in Abel 1201 ($z_{\rm lens} = 0.169$), providing the first lensing-based measurement of an SMBH mass with limits placed on both sides \citep{bib3:Nightingale23}. Additionally, \citet{bib3:Millon23} demonstrated the usefulness of strong lensing \textit{by} a quasar (SDSS J0919$+$2720, shown in \figref{fig3:lens_montage}) to measure its host galaxy mass to robustly probe the SMBH--host mass relation, where the SMBH mass was measured through conventional methods based on spectroscopic data.}

\refedit{For other kinds of SMBHs,} prospects for detecting \refedit{binary SMBHs} due to their lensing effects are discussed in \citet{bib3:Li12, bib3:Hezaveh15}. \refedit{Free-floating SMBHs, which could be of primordial origin or formed through co-evolution with their previous host galaxies, can be detected through the kinks they produce on razor-thin arcs with sub-mas width and resolved in radio observations \citep[][]{bib3:Banik19}. The detection (or the lack thereof) of such free-floating SMBHs can place constraints (or upper limits) on their mass density and fractional contribution to the dark matter.}

\subsection{Spiral galaxies} \label{sec3:spiral_galaxies}
Due to their lower mass and presence of a disk, spiral galaxies have a substantially lower lensing cross-section than elliptical ones \citep{bib3:Keeton98}. Only a handful of spiral galaxies lensing quasars have been discovered to date. The most well-known and the first example of such a system is the Einstein Cross (2237$+$0305), which is lensed by the bulge of a nearby spiral galaxy \citep{bib3:Huchra85}. Targeted searches for galaxies lensed by a spiral have been carried out, allowing the discovery of several dozens of systems \citep[][and reference therein]{bib3:Treu11}. Complications in studying those systems arise from the dust and the disk mass component. Correcting for the reddening by dust is needed to model extended lensed images, which rely on conserving the surface brightness between the lens and the source plane. On the other hand, the disk component yields strong discontinuities in the gravitational potential and needs to be explicitly modeled using, for example, an exponential mass density profile. Whereas disk-like features can yield flux anomalies in lensed quasars \citep{bib3:Hsueh16, bib3:Hsueh17, bib3:Gilman17}, they can be disentangled from the dust using multi-band data \citep{bib3:Moeller03}. Once these ingredients are accounted for, the combination of kinematics and lensing information can be used to break the degeneracy -- that exists in dynamical studies alone -- between the disk and halo components \citep{bib3:Maller00, bib3:Dutton11, bib3:Suyu12b}. %

\section{Applications in cosmology} \label{sec3:cosmology}

In this section, we present applications of galaxy-scale lens systems to measure cosmological parameters without using time-delay information. Measurement of the Hubble constant ($\Hzero$) and other cosmological parameters based on the time delays are discussed in detail in \chapho. 

Here, we briefly establish some necessary definitions for use in this section. More detailed explanations of the cosmological connection with strong lensing formalism are given in \chapintro.
The angular diameter distance %
between two objects at redshifts $z_1$ and $z_2$ (with $z_1<z_2$) for a flat universe is given by
\begin{equation}
    D_{\rm ang}(z_1, z_2)=\frac{c}{H_0 (1+z_2)} \int_{z_1}^{z_2} \frac{\mathrm{d}z'}{E(z')}, 
    \label{eq3:dist}
\end{equation}
where $E(z) \equiv H(z)/H_0$ is the dimensionless Friedman equation given by
\begin{equation}
    E(z) = \sqrt{{\Omegam} (1+z)^3 + {\Omegar} (1+z)^4 + {\Omega}_{\rm de} (1+z)^{3(1+w_{\rm de})}}.
    \label{eq3:wCDM}
\end{equation}
Here, ${\Omegam}$, ${\Omegar}$, and ${\Omega}_{\rm de}$ represent the matter, radiation, and dark energy density parameters, respectively, at $z=0$. The parameter $w_{\rm de}$ is the equation-of-state parameter of the dark energy given by $w_{\rm de} \equiv p_{\rm de}/\rho_{\rm de} c^2$, where $p_{\rm de}$ and $\rho_{\rm de}$ denote the pressure and density of the dark energy, respectively. In the $\Lambda$CDM model, $w_{\rm de} = -1$ is assumed, that is, the dark energy density stays constant through cosmic time. The $w$CDM model is one natural extension of the $\Lambda$CDM model, where $w_{\rm de} \neq -1$ is allowed; however, $w_{\rm de} < -1/3$ should still be satisfied to reproduce an accelerated Universe.%

In the following subsections, we discuss methods to estimate cosmological parameters (primarily, $\Omegam$, $\Omega_{\rm de}$, and $w_{\rm de}$) using multiple-source-plane lenses (\secref{sec3:double_source_plane}), %
using the stellar kinematics of strong lenses (\secref{sec3:single_source_cosmology}), and using galaxy--galaxy lensing statistics (\secref{sec3:lensing_statistics}). %

\begin{figure}
\centering
	\includegraphics[width=1\textwidth]{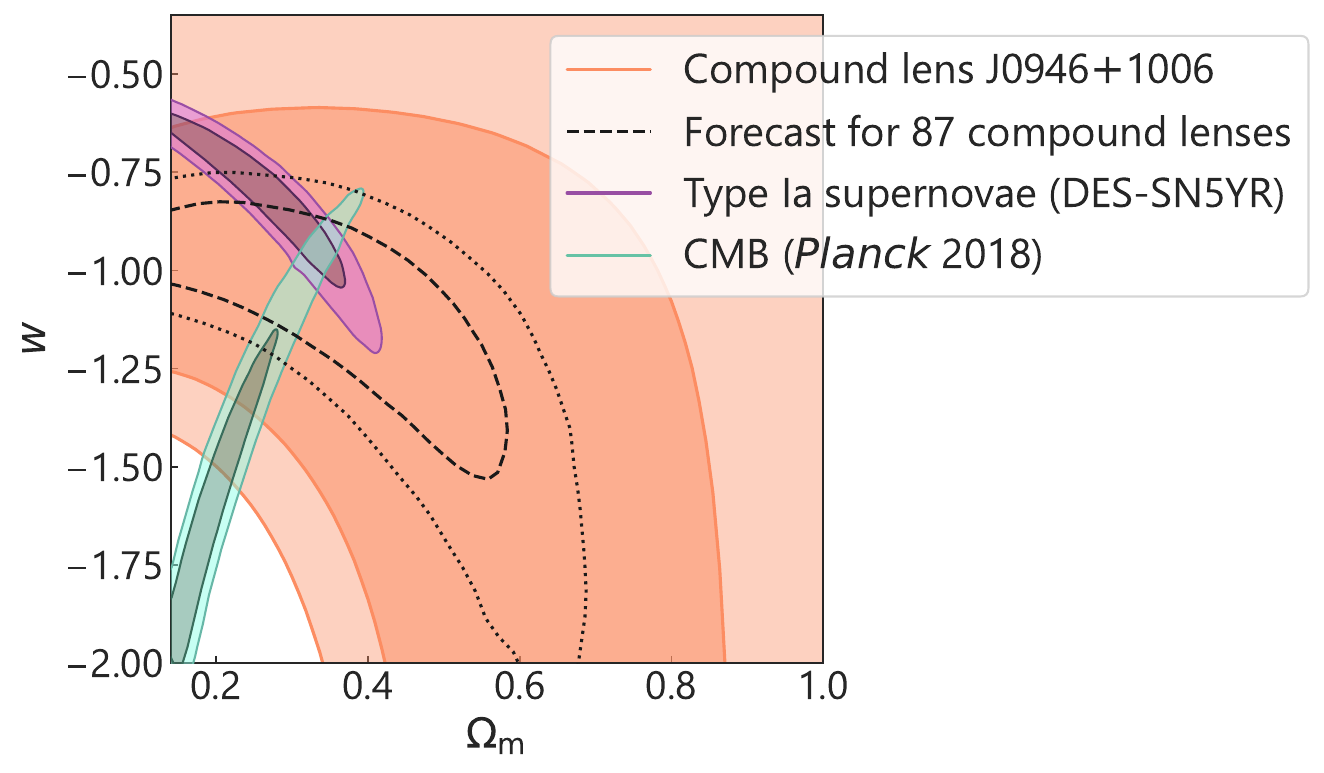}
	\caption{\label{fig3:double_source_plane_lens}
	\refedit{Cosmological parameters constrained from the compound lens system SDSS J0946$+$1006 (shown in \figref{fig3:double_ring}). The 68\% and 95\% credible regions on the $\Omega_{\rm m}$--$w$ plane from this compound plane system are shown in orange \citep{bib3:Collett14}. The black dashed and dotted contours illustrate the forecasted 68\% and 95\% credible regions, respectively, for a sample of 87 compound lenses to be discovered by the Rubin Observatory LSST \citep{bib3:Sharma23}. The purple and emerald contours are the constraints from the DES Type Ia supernovae sample \citep{bib3:DES24} and the CMB \citep{bib3:Planck20}, respectively, illustrating the complementarity of compound lenses to these probes.}
 }
\end{figure}

\subsection{Utilizing multiple sources at different redshifts} \label{sec3:double_source_plane}

Strong lens systems with multiple sources (i.e., compound lenses) at different redshifts can be used as cosmographic probes. Currently, only a tiny sample of galaxy-scale compound lenses are known %
\citep[e.g.,][]{bib3:Lewis02, bib3:Gavazzi08, bib3:Collett20}, but at the \textit{HST} snapshot depth they are expected to occur in about 1\% of galaxy-scale strong lenses \citep{bib3:Gavazzi08}. Indeed, if one were to stare deeply at any single plane lens, other sources would almost inevitably be discovered, as is spectacularly demonstrated by the discovery of a third source (a $z \approx 6$ Lyman-$\alpha$ emitter) behind the Jackpot lens in deep Multi Unit Spectroscopic Explorer (MUSE) data from the Very Large Telescope \citep{bib3:Collett20}. Forthcoming surveys are expected to discover $\mathcal{O}(1000)$ compound lenses.

Since the Einstein radius is a function of the lens mass and the cosmological distances, the ratio of Einstein radii in a compound lens with sources at two or more redshifts is independent of the mass \citep{bib3:Gavazzi08, bib3:Collett12}. In practice, the method also requires a complete understanding of the lens density profile and additional lensing by the source galaxies and other perturbing masses along the line of sight.

For a two-source-plane system with one primary lens, the lens equation can be written as 
\begin{equation}
	\begin{aligned}
	&\mathbf{y} = \mathbf{x} - \beta_{12} \balpha_{\rm d} (\mathbf{x}), \\
	&\mathbf{z} = \mathbf{x} - \balpha_{\rm d} (\mathbf{x}) - \balpha_{{\rm s}1} \left(\mathbf{x} - \beta_{12} \balpha_{\rm d} (\mathbf{x})\right),
	\end{aligned}
\end{equation}
where $\bf x$ are positions on the image plane, $\bf y$ and $\bf z$ are the unlensed positions of the first and second source, respectively, $\balpha_{\rm d}$ is the deflection caused by the primary lens, $\balpha_{{\rm s}1}$ is the deflection caused by the closer of the two sources, and $\beta_{ij}$ is the cosmological scaling factor
\begin{equation}
\beta_{ij} = \frac{D_{ij} \ds}{D_j D_{i \rm s}}.
\end{equation}
For realistic redshifts and cosmologies, $\beta_{ij}$ is sensitive to the matter density parameter $\Omegam$ and the equation-of-state parameter $w_{\rm de}$ but has no dependence on the Hubble constant $\Hzero$.

Amongst galaxy-scale compound lenses, only the Jackpot lens  \citep{bib3:Gavazzi08}, shown in \figref{fig3:double_ring}, has been used to precisely constrain cosmology since this system has a favorable redshift configuration -- other compound lenses have multiple sources at similar redshifts, thus having $\beta \approx 1$ regardless of the cosmology. \citet{bib3:Collett14} modeled the \textit{HST} imaging performing a pixellated reconstruction of both sources to make a 1.1\% measurement on $\beta^{-1}$. Converting this into constraints on the dark energy, this single compound lens with a cosmic microwave background (CMB) prior from \cite{bib3:Planck13XVI} constrains $w_{\rm de}$ to 0.2 precision. With hundreds of compound lenses expected in the Rubin Observatory Legacy Survey of Space and Time (LSST) and the \textit{Euclid} surveys, constraints on both $w_{\rm de}$ and its redshift derivative are expected to be \refedit{comparable} with established cosmological probes \refedit{\citep[\figref{fig3:double_source_plane_lens}; ][]{bib3:Sharma23}}.

\subsection{Utilizing stellar kinematics of single source lenses} \label{sec3:single_source_cosmology}

The enclosed projected mass inside the Einstein radius is independent of the mass profile \citep{bib3:Schneider92}. { Similarly, the speeds of the stars within a galaxy are sensitive to the total mass enclosed within their orbits. Whilst we cannot measure the speeds of individual stars in a lens, we can measure the velocity dispersion of the ensemble}. 
For an SIS lens with stars on isotropic orbits, these two quantities are related by
\begin{equation}
    \tae = 4 \pi  \left( \frac{\sigma_{\rm los}}{c} \right)^2  \frac{\dds}{\ds},
    \label{eq3:sis}
\end{equation}
where $\rm \sigma_{\rm los}$ is the velocity dispersion of the deflector. %
{\cite{bib3:Ofek03} estimated that deviations from isothermality and orbital isotropy can cause the observed velocity dispersion to differ from \eqref{eq3:sis} by up to 20\%.}

{\eqref{eq3:sis} becomes far more complicated if the density profile is not isothermal or the orbits of the stars are not isotropic, but the fundamental relationship remains that one can constrain the distance ratio $d^{\rm obs} \equiv \dds/\ds$ using the observed Einstein radius $\tae$ and observed velocity dispersion.}
Thus, a sample of single-source lenses with measured $\sigma_{\rm los}$ can be used to estimate cosmological parameters \citep{bib3:Grillo08} by maximizing the likelihood function
\begin{equation}
    \mathcal{L}\left(\mathbf{\Theta} \mid \mathcal{D} \right) \propto \exp \left[ -\frac{1}{2} \sum_{i=1}^{N_{\rm SL}} \frac{ \left[ d^{\rm th}_i\left(\zd, \zs; \bf{\Theta} \right) - d^{\rm obs}_i({\tae}_{,i}, \sigma_{{\rm los}, i})\right]^2 }{ (\delta d^{\rm{obs}}_i)^2} \right],
\end{equation}
where $\mathbf{\Theta}$ is {a set consisting of the free parameters in the assumed cosmological model and the free parameters that describe the density profile and anisotropy profile of the lenses}, $\mathcal{D}$ is the data, $N_{\rm SL}$ is the number of lens systems, and $\delta d_i^{\rm{obs}}$ is the uncertainty of each $d_i^{\rm obs}$, which depends on the $\sigma_{{\rm los}, i}$ and ${\tae}_{,i}$ uncertainties.

Similar to the multiple-source lens systems described in \secref{sec3:double_source_plane}, this method is also independent of $\Hzero$. \refedit{If the cosmological parameters and the lens population parameters are inferred simultaneously for a large sample of lenses (e.g., $N_{\rm lens}\sim$10~000) discovered by the Rubin Observatory LSST and \textit{Euclid}, it will be possible to achieve very competitive precision with other probes such as Type Ia supernovae and the CMB \citep{bib3:Li24}.} 

The primary systematics in this method {can potentially arise from these assumptions}: (i) the measured $\tae$ is independent of the choice of the lens mass profile \citep[e.g.,][]{bib3:Cao15}, and (ii) the measured line-of-sight velocity dispersion is equal to that for an SIS profile. Recent modeling methods provide robust $\tae$ measurement within a few percent regardless of the mass profile choice \citep{bib3:Birrer21b}. \citet{bib3:Treu06} argue that $\sigma_{\rm los}\simeq \sigma_{\rm SIS}$ for the lens elliptical galaxies with velocity dispersion in the range 200--300  km s$^{-1}$. These results are further confirmed by analyzing other samples \citep{bib3:Bolton06, bib3:Auger10b}. %

{Ultimately, the only way to move forward with lensing and dynamics as a precision cosmological probe is to simultaneously infer the astrophysical parameters of the lens population and the cosmological parameters. By building up a sample of many lenses it should be possible to investigate how the Einstein radius grows with source redshift regardless of the underlying density profile of strong lenses. Exploiting the fact that lenses at the same redshift can be expected to be somewhat self-similar, \citet{bib3:Li24} showed that with 10\,000 lenses, it is possible to disentangle lens population properties and cosmological parameters. These authors assumed that lenses have the same intrinsic scatter as \citet{bib3:Auger10b} found for the SLACS lenses. The method is fundamentally limited by how self-similar lenses are, and if their properties evolve with redshift.}


\subsection{Utilizing galaxy--galaxy lensing statistics} \label{sec3:lensing_statistics}

The statistics of strong lensing were initially expected to be powerful for probing the cosmological parameters \citep{bib3:Fukugita90}. At the most basic level, matter clusters can collapse to the densities required to form strong lenses, whereas the cosmological constant does not cluster and cannot form lenses. Therefore, it was expected that lensing rates should be suppressed for larger values of $\Omega_\Lambda$. Further information is contained in the Einstein radius distribution and the lens and source redshift distributions. In practice, this topic has fallen out of fashion due to the cosmological sensitivity being overwhelmed by astrophysical uncertainties of the unlensed source population, the lens discovery selection function, and the lensing properties of typical galaxies \citep[e.g.,][]{bib3:Mitchell05, bib3:Chae10}.

\section{Open problems and future outlook} \label{sec3:open_problems}

In this section, we discuss the current open problems that are expected to be tackled in this decade: the selection function (\secref{sec3:selection_functions}), the self-similarity assumption (\secref{sec3:self_similarity}), degeneracies in strong-lensing (\secref{sec3:lensing_degeneracy_open_problem}) and non-strong-lensing observables (\secref{sec3:non_lensing_degeneracy_open_problem}), and comparison with galaxy simulations (\secref{sec3:simulation_issues}). %
We also provide future outlooks on these issues whenever appropriate.

\subsection{Selection function} \label{sec3:selection_functions}

\begin{figure}
\centering
	\includegraphics[width=\textwidth]{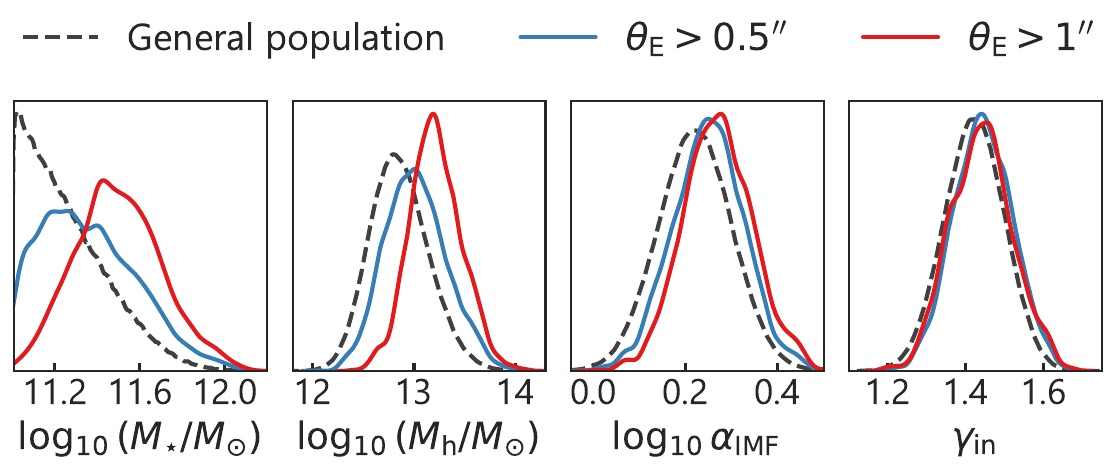}
	\caption{\label{fig3:selection_effect}
    \refedit{Impact of strong lensing selection function, estimated by \citet{bib3:Sonnenfeld23}, on various galaxy properties: from left to right, stellar mass $M_{\star}$, halo mass $M_{\rm h}$, IMF mismatch parameter $\alpha_{\rm IMF}$ with $\alpha_{\rm IMF} = 1$ corresponding to the Chabrier IMF, and the dark matter's inner logarithmic slope $\gamma_{\rm in}$. The grey dashed line shows the distribution of the general population of galaxies. The blue and red lines show the distributions of galaxy--galaxy strong lenses with $\tae > 0\,\arcsecf{5}$ and $\tae > 1\arcsec$, respectively. Strong lenses are expectedly biased toward more massive galaxies. The mean IMF mismatch parameter in strong lensing galaxies is biased by 10\% from the general population, and the mean inner logarithmic slope is biased by 5\%.
    }
    }
\end{figure}

Although strong lensing is a powerful probe to study galaxy properties, the lensing phenomenon is a rare occurrence requiring a serendipitous alignment of two line-of-sight objects separated by a large cosmological distance. Thus, samples of strong lensing galaxies inherently occupy a tiny fraction of the population of all galaxies. When strong lensing studies aim to infer properties of the general population of galaxies based on such a small fraction of galaxies, the lens sample's selection function must be considered. \refedit{The strong lensing samples are inherently biased towards lensing galaxies that are more massive and concentrated \citep{bib3:Mandelbaum09, bib3:Sonnenfeld23}. Although triaxial galaxies with the major axis more aligned along the line of sight also have larger lensing cross-sections, interestingly for a given mass and shape, the effect of viewing angle does not affect the selection function when averaged over \citep[][we note that this study only considered point sources]{bib3:Mandelbaum09}.} \refedit{ \citet{bib3:Sonnenfeld23} estimate that the mean of the IMF mismatch parameter $\alpha_{\rm IMF}$ measured from a sample of lens galaxies is only biased by 10\% and the mean of the inner slope of the dark matter by 5\% (\figref{fig3:selection_effect}). These bias levels are dependent on the completeness in the Einstein radius distribution of the lens sample but independent of the source properties, with the galaxy--galaxy lenses and galaxy--quasar lenses having the same levels of bias.}

\refedit{The lens samples to date often had highly complex selection functions, largely due to the selection or discovery procedure being highly tuned to maximize the number of discovered systems. As a result, treatment of selection function on actual lens samples has been rare except for a handful of studies \citep{bib3:Arneson12, bib3:Sonnenfeld15, bib3:Sonnenfeld19b}. For the ideal case of a known selection function, \citet{bib3:Sonnenfeld22} provides the formally correct solution to account for selection effects in the lens samples. However, several technical challenges still remain to implement this formal solution on real samples, for example, the requirement to well characterize the lens-detection efficiency of the survey and also the efficiency in obtaining follow-up spectroscopy if that was taken into account for sample selection. The treatment of the selection effect with parametric functions by \citet{bib3:Sonnenfeld19b} can be considered as an approximation to this formal solution. To keep the selection function easily treatable, it can be advisable to pre-emptively mitigate the complexity of the selection procedure. Alternatively, it can be possible to form a subsample of lens systems that has a well-characterized selection function from a much larger sample of discovered systems, for example, those discovered by current and future surveys such as the \textit{Euclid}, the Rubin Observatory LSST, and the \textit{Roman Space Telescope} \citep{bib3:Collett15, bib3:Sonnenfeld22}.}

\subsection{Assumption of self-similarity} \label{sec3:self_similarity}
When hierarchically inferring population properties of lenses %
by combining multiple strong lensing systems (\secref{sec3:hierarchical_framework}), it is crucial to describe and quantify potential differences between subsets of the considered sample for accurate population-level inference. Differences in the population may arise from the sample selection due to search criteria and techniques (\secref{sec3:selection_functions}) or intrinsic differences in the sources. Such differences in secondary selection might impact and bias population-level constraints when assuming that two different samples can be described with an identical underlying population. For example, \cite{bib3:Birrer20} present a hierarchical analysis under the assumptions that the same population level parameters describe the SLACS and the Time-Delay COSMOgraphy (TDCOSMO) lenses to constrain the mass density profiles of the time-delay lenses better. There are different ways to mitigate such assumptions. The first is to select a purified sample of lenses as self-similar as possible when performing hierarchical analyses, also suggested in \secref{sec3:selection_functions}.
However, although the currently small number of known lenses will increase through future surveys and enable this approach, it is still unclear in which parameter space these lenses can be considered self-similar and whether there are important latent variables to consider.  
The second way is to model and describe differences between populations on a first-principle level, folding in differential selection effects and other aspects into the analysis (as stated in \secref{sec3:selection_functions}). %

\subsection{Degeneracies in strong-lensing observables} \label{sec3:lensing_degeneracy_open_problem}

The lensing data's ability to constrain the lens's radial structure, particularly in disentangling dark and luminous mass components, strongly depends on the data quality. As discussed in \chapintro, the modeling of point-source astrometry primarily encodes information on the lens's quadrupole moment. Still, it provides limited constraints on the monopole (the total mass). This, however, does not mean that any monopole model combined with a quadrupole component will accurately reproduce a set of lensed image positions. For some choice of the monopole, extreme or unphysical values of the quadrupole may be required, naturally excluding some mathematically correct solutions. This explains why, for example, a single component constant mass-to-light ratio model generally yields large shear amplitudes to reproduce the observed astrometry of lensed systems to high accuracy \citep[e.g.,][]{bib3:Sluse12}.

While extended lensed images may provide detailed azimuthal information and allow one to constrain the ratio of radial magnifications at different galactocentric distances, their effective constraint on the density profile remains sensitive to the MSD \citep[e.g.,][]{bib3:Sonnenfeld18}. While the MSD provides mathematically large leverage to modify the results, its impact may remain in practice generally small, with a typical change on the total density profile that may not exceed a few tens percent. The prior on the mass profile (i.e., choice of mass distribution families or a free-form model with some regularization) may further limit the impact of degeneracies. One may, however, need to be careful when choosing a mass distribution, as a model that is too rigid compared to the true mass density may yield biased posteriors or underestimated parameter uncertainties \citep[e.g.,][]{bib3:Sonnenfeld18, bib3:Kochanek21}. Finally, it is important to note that point images with measured time delays limit the impact of degeneracies in lensing-only observables for an assumed cosmology \citep{bib3:Saha01, bib3:Kochanek02}.

\subsection{Degeneracies in non-strong-lensing observables} \label{sec3:non_lensing_degeneracy_open_problem}

The modeling of stellar kinematics data requires an assumption on the anisotropy profile of stellar orbits. Integrated velocity dispersions obtained from single-slit spectra cannot constrain the anisotropy profile, which leads to the so-called mass--anisotropy degeneracy \citep{bib3:Treu02b} --- typically adopted anisotropy profiles are either isotropic or the Osipkov--Merritt profile \citep{bib3:Osipkov79, bib3:Merritt85, bib3:Merritt85b}. Whereas the isotropy assumption does not entail any free parameter, the Osipkov--Merritt profile depends on a scale radius. Due to the mass--anisotropy degeneracy, the posterior of the anisotropy scale radius is dominated by the adopted prior \citep[e.g.,][]{bib3:Shajib18}. Particular choices of the anisotropy profile and the associated prior may lead to systematic differences between studies involving strong lensing and kinematics data. For example, \citet{bib3:Sonnenfeld18c} find the existence of a mass-to-light ratio gradient in the SLACS lenses assuming isotropic orbits, whereas \citet{bib3:Shajib21} find consistency with a constant mass-to-light ratio assuming the Osipkov--Merritt anisotropy profile for a subsample of SLACS. \refedit{Additionally, the unknown 3D structures of the mass distribution and the tracer distribution are also potential sources of systematics \citep{bib3:Cappellari08}.}

 Spatially resolved velocity dispersion measurements can better constrain the anisotropy profile by breaking the mass--anisotropy degeneracy. However, spatially resolved kinematics data from IFU spectroscopy are more expensive than an integrated measurement from long-slit spectroscopy. Thus, usage of such data in lensing studies has been limited \citep[see][for example of IFU data being combined with strong lensing]{bib3:vandeVen10, bib3:Barnabe11, bib3:Shajib23}.
 
 \subsection{Comparison with galaxy simulations} \label{sec3:simulation_issues}

The past decade has seen significant progress in understanding galaxies' structure and formation. Within the $\Lambda$CDM paradigm, there is
general agreement regarding the gravitational \refedit{aspect} of galaxy formation.
The `gastrophysics' is less well understood and requires subgrid
models to simulate, but still, the galaxies formed in simulations like
Illustris \citep{bib3:Vogelsberger14}, FIRE
\citep{bib3:Hopkins14}, EAGLE \citep{bib3:Crain15}, \refedit{IllustrisTNG \citep{bib3:Nelson}} are
much more credible than previous generations of simulated galaxies.
There are also equilibrium galaxy models, including stars, gas, and
dark matter, of which the AGAMA \citep{bib3:Vasiliev19}
simulations are arguably the most sophisticated.

Despite the great advances in the fidelity of the simulations, there have been discrepancies between the simulated predictions and the observed properties of galaxies. In particular, simulations have not been successful yet in reproducing the observed distributions of the logarithmic slope $\gammapl$ and the dark matter fraction $f_{\rm dm}$ simultaneously. A no-feedback or weak feedback prescription was required in some of the simulations to reproduce the $\gammapl$ distribution, which, however, led to underestimating $f_{\rm dm}$ compared to the observations due to overestimating the star formation efficiency \citep[e.g.,][]{bib3:Naab07, bib3:Duffy10, bib3:Johansson09}. Similarly, matching the $f_{\rm dm}$ distribution required strong feedback prescriptions, but these produce too shallow $\gammapl$ compared to the observed distribution. More recently, the IllustrisTNG simulation reproduced a $f_{\rm dm}$--$\gammapl$ distribution that is consistent with the strong lensing observations if the stellar IMF corresponds to the Salpeter IMF \citep{bib3:Wang20, bib3:Shajib21}. In contrast, \cite{bib3:Mukherjee22}, find that the EAGLE and SLACS  $f_{\rm dm}$--$\gammapl$ distributions agree while using a Chabrier IMF, supporting the important role played by feedback and sub-grid physics in reproducing this relation. 

\begin{figure*}[!]
	\includegraphics[width=1\textwidth]{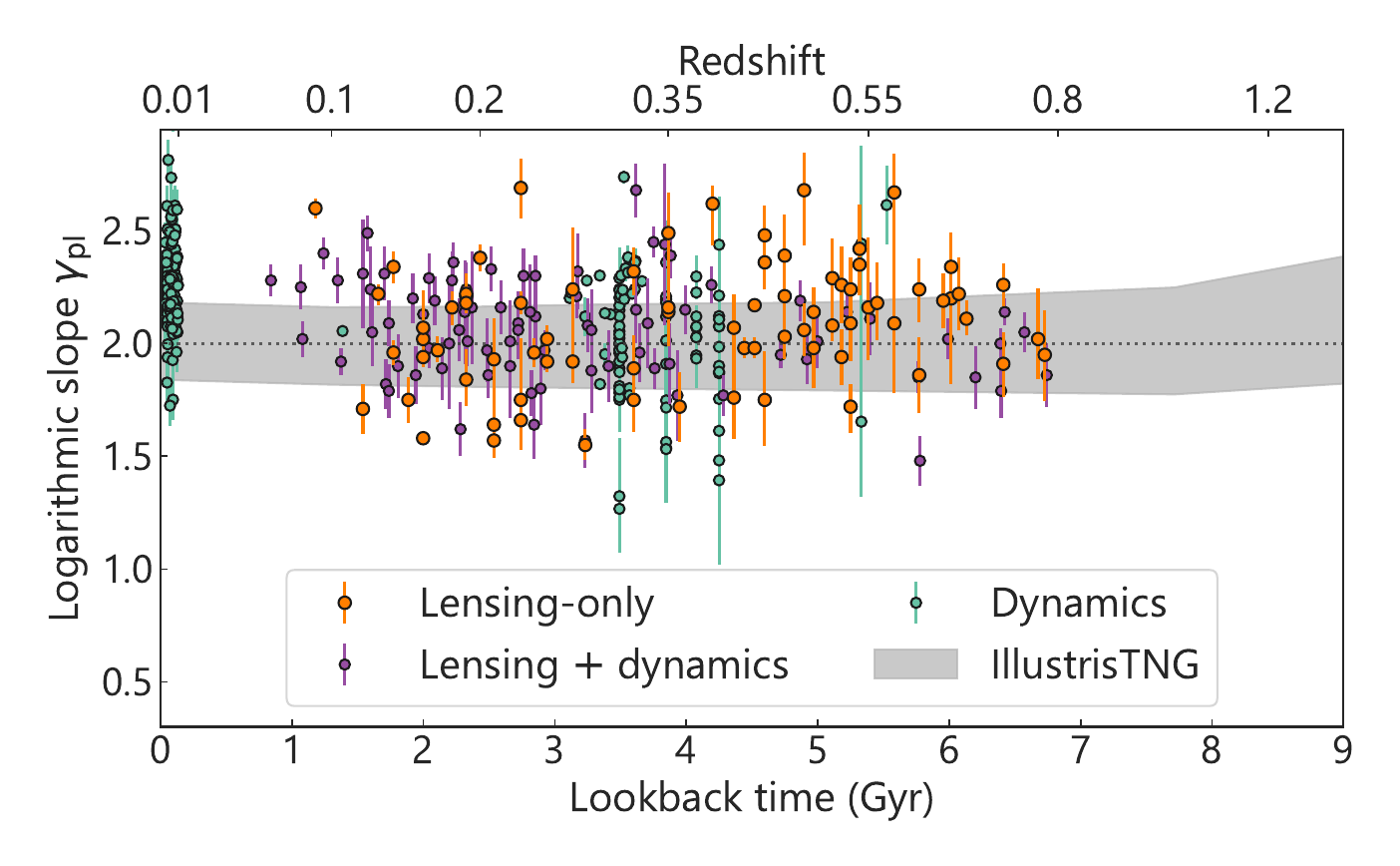}
	\caption{ \label{fig3:gamma_redshift_evolution}
	\refedit{Comparison of measured logarithmic slopes $\gammapl$ of the total density profile at different redshifts between lensing-only measurements (orange points), lensing--dynamics measurements (purple points), dynamical measurements (emerald points), and the IllustrisTNG simulation \citep[grey shaded region; ][]{bib3:Wang20}. The lensing-only measurements are from the SLACS, SL2S, and BELLS samples \citep{bib3:Tan24}. The lensing--dynamics measurements are from the SLACS and SL2S samples \citep{bib3:Auger10b, bib3:Sonnenfeld13b}. The dynamical measurements are from the ATLAS$^{\rm 3D}$, the Frontier Fields, and the Middle Ages Galaxy Properties with Integral field spectroscopy (MAGPI) surveys \citep{bib3:Poci17, bib3:Derkenne21, bib3:Derkenne23}. The horizontal dotted line traces $\gammapl=2$, the isothermal case.} %
	}
\end{figure*}

Furthermore, there has been an apparent tension in the redshift evolution of the logarithmic slope $\gammapl$ between observations and simulations.
Strong lensing observations report a steepening of $\gammapl$ with decreasing redshift at $z < 1$ \citep{bib3:Ruff11, bib3:Sonnenfeld13b}. Such a steepening would require dissipative processes through wet mergers along the evolutionary track of elliptical galaxies. Simulations instead find no evidence for redshift evolution of $\gammapl$ below $z = 1$, or a slightly shallowing trend in $\gammapl$ with decreasing redshift \citep[see \figref{fig3:gamma_redshift_evolution},][]{bib3:Xu17, bib3:Remus17, bib3:Wang20}. \refedit{Strong lensing selection effects could be a potential source of this discrepancy \citep{bib3:Sonnenfeld15}. Moreover,} this tension vanishes if the same strong lensing analysis is applied to the simulated galaxies from Illustris and Magneticum to extract $\gammapl$ by combining lensing and kinematic information \citep{bib3:Xu17, bib3:Remus17}. Therefore, the modeling systematics in the joint analysis of lensing and dynamical observables cannot be ruled out as \refedit{another potential} source of the above tension. \refedit{For an example of modeling systematic, if the true mass distribution in lensing elliptical galaxies is not an accurate power law, then lensed images are formed at different galactocentric radii as the lens and source redshifts vary \citep[for a relevant test of systematic, see][]{bib3:Gomer22}. In that case, minor deviations from the adopted power-law model may mimic an (absence of) evolution of the galaxy's total density profile with redshift. To account for this effect, \citet{bib3:Dutton14} suggest using a mass-weighted slope.} %

\refedit{An alternative approach to compare observed galaxy properties to theoretical models is the semi-empirical one adopted by \citet{bib3:Shankar17, bib3:Shankar18}. These authors find their semi-empirical model for massive elliptical galaxies to be consistent with un-contracted NFW halos and the Salpeter IMF, corroborating with the majority of the previous literature (see \secref{sec3:dark_structure} and \secref{sec3:imf_results}). These authors also investigate the redshift dependence of the logarithmic slope $\gammapl$ mentioned above and find that a redshift dependence of the S\'ersic index is necessary to explain it. In contrast, the selection function does not contribute to producing the redshift-dependent trend in $\gammapl$.}

The specific importance of various baryonic processes in the evolution of ellipticals is still an open question. The simulations would require further fine-tuning to consistently reproduce all of their observable properties. However, modeling systematics and the selection function must be appropriately considered for accurate comparison between simulations and observations. The first results from a statistical framework jointly considering simulations and observations to infer the galaxy evolution scenario indicate that AGN feedback is essential \citep{bib3:Denzel21}, but application to a large sample of lenses is required.

\refedit{An additional usage of simulated galaxies is to take them as the deflector galaxies in synthesizing strong-lensing observables for testing and validating the assumptions made in the simply parametrized mass models \citep[e.g.,][]{bib3:Enzi20, bib3:Ding21}. These investigations often find the simple power-law parametrization to be inadequate to accurately describe the data \citep[e.g.,][]{bib3:He23}. However, numerical inadequacies in synthesizing the strong-lensing observables can also potentially hamper such investigations \citep{bib3:Vandevyvere20}. Alternatively, simply parametrized lens models can also be tested based on independent empirical observations, for example, using high-resolution imaging of nearby ellipticals \citep{bib3:Gilman17}, or using dynamical models from highly resolved IFU spectroscopy of ellipticals \citep[e.g.,][]{bib3:Cao22, bib3:Poci22}. In the future, exquisite high-resolution imaging from the \textit{JWST} or extremely large telescopes, or advanced dynamical models, such as Schwarzschild models \citep[extending the original method of][]{bib3:Schwarzschild79}, will provide powerful means to carry out these important validation tests.}

\section{Concluding remarks} \label{sec3:conclusion}

In this \thischapter, we have provided a review of the applications of galaxy-scale strong lensing in astrophysics and cosmology. Inevitably, some special sub-topics within the field of galaxy-scale strong lensing have evolved into proper research fields, having acquired a methodology and literature extensive enough to warrant a dedicated review. These are: detecting dark matter substructures and linking their properties to the dark matter particle, and measuring $\Hzero$ and other cosmological parameters through time delays, examined in \chapdm\ and \chapho, respectively.

We started with a brief historical overview in \secref{sec3:historical_results}. Then, in \secref{sec3:observables_and_methods}, we have discussed both strong-lensing and complementary non-strong-lensing observables and methodologies to model and extract meaningful results from such data. The most available and informative data for galaxy-scale strong lenses come from imaging. We reviewed the most common modeling methods found in the literature to model such data and constrain galaxy properties from lensed arcs, and in some cases with the inclusion of multiple images of a point-like source. Next, we reviewed the main scientific results from the literature on the astrophysics of galaxies in \secref{sec3:science_applications} and on cosmology in \secref{sec3:cosmology}. We then discussed the currently open questions and provided future outlooks in \secref{sec3:open_problems}.

The open questions presented in \secref{sec3:open_problems} provide exciting opportunities for the near future. Several large-area sky surveys -- namely the Rubin, \textit{Euclid}, and \textit{Roman} observatories -- will discover thousands of new galaxy-scale lensing systems. %
These treasure troves of data will provide the necessary statistical power to shed light on the open questions on galaxy evolution and cosmology.

\begin{acknowledgements}
The authors thank the anonymous referees for helpful comments that improved the content and quality of the manuscript. The authors also thank Caro Derkenne and Nandini Sahu for providing a compilation of measurements from the literature, which was used in \figref{fig3:gamma_redshift_evolution}.
The authors thank the International Space Science Institute in Bern (ISSI) for their hospitality and the conveners for organizing the stimulating workshop on ``Strong Gravitational Lensing''.
Support for this work was provided by NASA through the NASA Hubble Fellowship grant HST-HF2-51492 awarded to AJS by the Space Telescope Science Institute, which is operated by the Association of Universities for Research in Astronomy, Inc., for NASA, under contract NAS5-26555.
This work was also supported by the U.S. Department of Energy (DOE) Office of Science Distinguished Scientist Fellow Program.
GV has received funding from the European Union's Horizon 2020 research and innovation program under the Marie Sk\l{}odowska-Curie grant agreement No 897124.
This research was made possible by the generosity of Eric and Wendy Schmidt by recommendation of the Schmidt Futures program.
VM acknowledges partial support from Centro de Astrof\'{\i}sica de Valpara\'{\i}so and project "Fortalecimiento del Sistema de Investigaci\'on e Innovaci\'on de la Universidad de Valpara\'{\i}so" (UVA20993).

This article made use of \textsc{lenstronomy} \citep{bib3:Birrer18, bib3:Birrer21},  \textsc{astropy} \citep{bib3:AstropyCollaboration13, bib3:AstropyCollaboration18, bib3:Astropy22}, \textsc{numpy} \citep{bib3:Oliphant15}, \textsc{scipy} \citep{bib3:Jones01}, \textsc{matplotlib} \citep{bib3:Hunter07}, \textsc{seaborn} \citep{bib3:Waskom14}, and \textsc{ChainConsumer} \citep{bib3:Hinton16}.

\end{acknowledgements}

\section*{Declarations}

\textbf{Competing Interests} The authors have no conflicts of interest to report.

\bibliographystyle{aps-nameyear}      %
\bibliography{references}                %

\end{document}